%% file: arxiv-V1.tex
\documentclass[11pt,a4,twoside]{article}

\usepackage{./styling/style-main}

\title{A critical look at the electroweak phase transition}

\author{Andreas~Ekstedt\thanks{andreas.ekstedt@ipnp.troja.mff.cuni.cz}\textsuperscript{~~,\,a}}
\author{Johan~L\"ofgren\thanks{johan.lofgren@physics.uu.se}\textsuperscript{~~,\,b}}
\affil{a:~Institute of Particle and Nuclear Physics, Charles University \\
b:~Department of Physics and Astronomy, Uppsala~University, Box~516, SE-751~20~Uppsala, Sweden}

\date{\today}

\begin{document}
	\maketitle
	\thispagestyle{plain}

	\begin{abstract}
    The electroweak phase transition broke the electroweak symmetry.
    Perturbative methods used to calculate observables related to this phase transition suffer from severe problems such as gauge dependence, infrared divergences, and a breakdown of perturbation theory. In this paper we develop robust perturbative tools for dealing with phase transitions. We argue that gauge and infrared problems are absent in a consistent power-counting. We calculate the finite temperature effective potential to two loops for general gauge-fixing parameters in a generic model. We demonstrate gauge invariance, and perform numerical calculations for the Standard Model in Fermi gauge.
	\end{abstract}
\graphicspath{{Plots/}}
	\input{./tex/introduction}
  \input{./tex/powers}
  \input{./tex/phases}
  \input{./tex/examples}
  \input{./tex/results}
  \input{./tex/discussion}
  \appendix
	\input{./tex/integrals}

	\let\enquote\letmacundefined 
	\bibliographystyle{bibstyle}
	{\bibFont\RaggedRight
	\bibliography{Bibliography}%
	}%
\end{document}

%% file: tex/introduction.tex
\section{Introduction}
The electroweak symmetry appears exact in the early universe, but not in our current day and age.
As the universe expands and cools down the Higgs field develops a vacuum-expectation-value (VeV)\te breaking the symmetry. There is a phase transition from a symmetric to a broken phase.

Although well established in the Standard Model, the \textit{electroweak phase transition} remains elusive. It is, as yet, unknown when and how the transition took place; if it was violent, or calm; first-order, or continuous. Continuous transitions are rather innocuous compared to their first-order cousins.
Indeed, a first-order phase transition is a turbulent and violent affair that likely has far-reaching consequences for the universe's development. Such transitions are part and parcel for understanding the observed matter-antimatter asymmetry~\cite{Bennett:2003bz}. Furthermore, gravitational waves from a strong phase transition reverberate throughout the universe and might be picked up by next-generation experiments~\cite{Caprini:2019egz}. Describing these phenomena goes hand-in-hand with understanding phase transitions.

Both perturbative and lattice methods accomplish this task; both methods with their fair share of virtues and vices.

On the perturbative side, the name of the game is the effective action.
Calculations are carried out with Feynman diagrams\te{}both quantum and temperature effects are included in this approach. But problems loom around the corner.

It is unclear if perturbative methods can at all be trusted~\cite{Morrissey:2012db,Senaha:2020mop}. There are ample issues related to gauge invariance~\cite{Patel:2011th}; appearance of infrared divergences~\cite{Andreassen:2014eha}; not to mention a breakdown of perturbation theory itself~\cite{Kapusta}.
By charging headlong one might even erroneously conclude that first-order phase transitions disappear for some gauge choices. And the perturbative expansion is likewise dicey. It can, and does, break down. Consider a scalar theory with quartic coupling $\lambda$. Finite temperature calculations include diagrams $d_N$, at loop order $N$, scaling as $d_N \sim T (\lambda T^{2})^{N-1}$. So loops are not suppressed for large temperatures~\cite{Arnold:1992rz}: $d_N / d_{N-1} \sim \lambda T^2\sim 1~\text{for}~ T^2\sim \lambda^{-1}$.
What's more, the phase transition occurs at these very temperatures\te{}as we discuss in section~\ref{ss:modCounts}.
So perturbation theory slowly but surely breaks down. This is not surprising in itself. After all, phase transitions occur when loop corrections overpower tree-level terms, which calls the loop expansion into question.

Yet it is long known how to alleviate these problems~\cite{Arnold:1992rz,Kapusta,Laine:2016hma}: a resummation is needed. As particles interact with the thermal bath they receive a thermal mass~\cite{Kapusta}\te{}a reorganization of the perturbative expansion around this effective mass improves convergence.
Although the need for resummation is established, there's no consensus on the implementation. There are a number of conflicting strategies~\cite{Parwani:1991gq,Patel:2011th,Curtin:2016urg,Laine:2017hdk,Arnold:1992rz}. More often than not resummations are gauge dependent.

But perturbation theory is not the only avenue. Indeed, lattice calculations are quite good at describing phase transitions. Though not without their own share of issues. Lattice simulations are resource expensive\te large parameter scans are, as yet, unfeasible. Still, lattice is well-suited at studying single models. The Standard Model's phase structure has indeed been investigated via lattice calculations~\cite{Rummukainen:1998jf,Gurtler:1997hr,Kajantie:1996mn,Karsch:1996yh}.

Perturbative calculations are on the other hand computationally cheap; so they are felicitous for studying complicated models. In this paper we propose a gauge invariant resummation. We develop robust perturbative techniques for describing phase transitions. These techniques are gauge invariant and aren't stymied by IR-divergences. Sticking to a strict power counting scheme is integral; a proper perturbative expansion\te with powers correctly accounted for\te is inherently gauge invariant. Our method is an amalgamation of (i) the early work of Arnold and Espinosa~\cite{Arnold:1992rz}, and (ii) the gauge invariant methods developed by Laine~\cite{Laine:1994zq} and emphasized by Patel and Ramsey-Musolf~\cite{Patel:2011th}. We restrict ourselves to observables at the critical temperature in this paper, and leave tunneling for the future.

%% file: tex/powers.tex
\section{The powers of perturbation theory}\label{sec:powers}
Perturbative calculations are organized in powers of a small quantity. This  might be a collection of couplings, or a ratio of energy scales. All terms must, at a given order, be included. The consequences of forgetting terms are dire\te{}including gauge dependence and exasperating divergences. The same holds when calculating the effective potential.

This section discusses subtleties and dangers of perturbative expansions. We introduce a systematic way to treat the breakdown of the "naive" loop expansion. We apply these considerations to the effective potential and show how and why a proper power counting is useful. The issues and concepts we discuss are well known, but we introduce our own notation.

\subsection{Power and loop counting}
In order to illustrate how a perturbative expansion might break down, and how it might be fixed, some terminology is in order. For example, in a standard loop expansion an observable $A$ is typically expanded as
\begin{equation}
  A = A_0 + \kappa A_1 +\kappa^2 A_2 +\mathellipsis
\end{equation}
with $\kappa$ denoting the number of loops.

Yet all is not fine and dandy. For if a coupling is large, the expansion might not be justified at all. And there are situations where calculations\te even in weakly coupled theories\te are not organized in loop powers.

So we better make a clear distinction between power- and loop-counting. To that end, introduce a new power counting parameter that better represents the actual sizes of terms: $\hb$. As a side-note, $\hb$ is not related to the reduced Planck constant which goes by the same symbol. We choose $\hb$ as the power counting parameter only to be congruous with earlier papers~\cite{Laine:1994zq, Patel:2011th}.

If the loop expansion is applicable, $\hbar$ and $\kappa$ are equivalent:
\begin{equation}
  A = A_0 + \hbar\kappa A_1 +\hbar^2\kappa^2 A_2 +\mathellipsis
\end{equation}
However, there might be terms in $A_n$ scaling with negative powers of $\hbar$. Consider a toy example, where $A_n=a_n+b_n / \hbar^{n-1}$  if $n \geq 2$. The expansion is
\begin{equation}\label{eq:toyexample}
    A = A_0 +\hbar \left( \kappa A_1 + \kappa^2 b_2 + \kappa^3 b_3 + \mathellipsis \right) + \hbar^2 \kappa^2 a_2 + \hbar^3 \kappa^3 a_3 +\mathellipsis,
\end{equation}
and diagrams from all loop orders are intertwined. If this is the case, a resummation is appropriate.

These ideas can be made lucid through a few examples.

\subsection{Gauging the problem} \label{ss:gaugeing}
The effective potential is in perturbation theory calculated order-by-order according to some power counting scheme, with $\hb$ denoting the aforementioned power:
\begin{equation}
V(\phi)=V_0(\phi)+\hb V_1(\phi)+\hb^2 V_2(\phi)+\mathellipsis
\end{equation}
The idea is to find the global minimum $\phim$, which then gives the physical energy density: $\Vmin \define V(\phim)$. The "standard" approach finds $\phim$ by minimizing $V(\phi)$ numerically. But this procedure is problematic and gives a gauge dependent $\Vmin$\te{}which doesn't make sense for a physical observable.

The effective potential at an arbitrary field value is not a physical observable, which the Nielsen identity makes glaringly clear~\cite{NIELSEN1975173},
\begin{equation}
\xi \partial_\xi V(\phi,\xi)+\mathcal{C}(\phi,\xi)\partial_\phi V(\phi,\xi)=0;
\end{equation}
$\xi$ is here a gauge-fixing parameter and $\mathcal{C}(\phi,\xi)$ is a calculable function known as the Nielsen coefficient. This is a non-perturbative statement describing the effective potential's gauge dependence. The equation suggests that $\Vmin$ is gauge invariant, but that $\phim$ necessarily depends on $\xi$. There needs to be a delicate cancellation between the gauge dependence of $\phim$ and $V$ for $\Vmin$ to be gauge invariant.

Why is it then not sufficient to minimize the potential numerically? The devil is in the details of perturbation theory, and a fiery analogy might be appropriate. Consider a particle's pole mass as calculated in perturbation theory,
\begin{equation}
  m_P^2=m^2+\hb \Pi_1(m_P^2)+\hb^2 \Pi_2(m_P^2)+\mathellipsis
\end{equation}
This is an implicit equation for $m_P^2$. Solve it by further expanding $m_P^2$ on the right-hand side, according to
\begin{equation}\label{eq:mP1}
  m_P^2=m_0^2+\hbar m_1^2+\mathellipsis,
\end{equation}
which gives the well-known result
\begin{equation}\label{eq:mP2}
  m_P^2=m^2+\hb \Pi_1(m^2)+\hb^2 \left[m_1^2\partial_{m^2}\Pi_1(m^2)+\Pi_2(m^2)\right]+\mathellipsis
\end{equation}
Comparing the two equations~\eqref{eq:mP1} and~\eqref{eq:mP2} order-by-order in $\hbar$, we deduce $m_0^2=m^2$ and $m_1^2=\Pi_1(m^2)$.
Likewise, as emphasized in~\cite{Patel:2011th},  $\phim$ must in turn be found order-by-order in $\hb$,
\begin{equation}
  \phim=\phi_0+\hb \phi_1+\mathellipsis
\end{equation}
Solving $\partial_\phi V(\phi)=0$ order-by-order in $\hb$ gives
\begin{align}
& \left.\partial_\phi\left[V_0+\hb V_1+\ldots\right]\right|_{\phim=\phi_0+\hb \phi_1+\ldots}=0
\\&\nonumber
\\ \implies\ordo{}(\hb^0):&\hspace{7.44em} \left.\partial_\phi V_0\right|_{\phi_0}=0,~
\\\implies\ordo{}(\hb^1):&\hspace{2em} \left.\left(\phi_1 \partial_\phi^2 V_0+\partial_\phi V_1\right)\right|_{\phi_0}=0.
\\&\hspace*{3.65cm}\vdots\nonumber
\end{align}

The minimum can then be plugged into the effective potential to give the physical and gauge independent energy density
\begin{align}
\Vmin&=\left[V_0+\hb V_1+\hb^2 V_2+\mathellipsis\right]\big|_{\phim=\phi_0+\hb \phi_1+\mathellipsis}\nonumber
\\&=V_0\big|_{\phi_0}+\hb V_1\big|_{\phi_0}+\hb^2 \left(V_2-\frac{1}{2}\phi_1^2 \partial^2 V_0\right)\Big|_{\phi_0}+\mathellipsis
\end{align}
Notice how all terms are expressed at $\phi_0$. This is expected since the expansion is organized around the leading order value.

Although consistent power-counting schemes are gauge independent, the  appropriate counting is not determined \emph{a priori}.
\subsection{Examples of modified power countings}\label{ss:modCounts}
\subsubsection{Coleman-Weinberg}\label{ss:CW}
A well known application of the effective potential is due to S. Coleman and E. Weinberg, where they establish the mechanism of quantum-generated spontaneous symmetry breaking~\cite{Coleman}. They considered scalar electrodynamics without a scalar mass term:
\begin{align}
\label{equation:ColemanWeinbergModel}
V_0=\frac{\lam}{4!}\phi^4.
\end{align}
There is no symmetry breaking at tree-level. But the symmetry can be still be broken by quantum effects\te{}the Coleman-Weinberg mechanism.

Explicitly, expand the potential as usual
\begin{equation}
V = V_0 +\hbar V_1 +\hbar^2 V_2+\mathellipsis,
\end{equation}
sub-leading corrections are given by scalar and photon loops
\begin{equation}
V_1 =\frac{3}{4} e^4 \phi^4 \left(\log\left[\frac{e^2 \phi^2}{Q^2}\right]-\frac{5}{6}\right)+\mathcal{O}\left(\lambda^2\right).
\end{equation}
How can the symmetry be broken by quantum corrections, which, after all, are suppressed in $\hb$? For this to happen, the 1-loop terms must compete with the tree-level terms. This indicates that $\lam$ must be small, $\lam \sim e^4$, compared to the standard (loop) counting $\lam \sim e^2$. This is accounted for by systematically counting lambda as $\hb$: $\lambda\rightarrow \hb \lambda$, implying
\begin{equation}
V = \hbar \left(\frac{\lam}{4!}\phi^4 + \kappa\frac{3}{4} e^4 \phi^4 \left(\log\left[\frac{e^2 \phi^2}{Q^2}\right]-\frac{5}{6}\right) \right)+\ordo(\hbar^2).
\end{equation}
Spontaneous symmetry breaking is now possible. Driven by a quantum effects.

But the situation is peculiar at higher orders. In Fermi gauge there are diagrams $d_N$ at $N $ loops scaling as $d_N \sim  e^{4 N} / \lambda^{N}$. The new power-counting $ \lambda \rightarrow \hbar \lambda\sim \hb e^4$ indicates that these terms are all of the same order. All of these terms must be included\te they must be resummed.
The authors in~\cite{Andreassen:2014eha} showed how this resummation in concert with the $\hb$-expansion gives a gauge invariant result. The resummation grabs the relevant terms from each loop order and organizes them in a gauge-invariant manner.

\subsubsection{Finite temperature effective potential}\label{ss:finiteT}
At finite temperature each propagator carries both three-dimensional momentum $\vec{p}$ and a Matsubara mode $p_0=2 \pi n T$. Bosonic tree-level propagators are of the form
\begin{equation}
  D(\vec{p},X)\propto\sum_{n=-\infty}^\infty\frac{1}{(2\pi n T)^2+\vec{p}^2+X+i\epsilon}.
\end{equation}
And for fermions the tree-level propagator is of the form
\begin{equation}
  D(\vec{p},X)\propto\sum_{n=-\infty}^\infty\frac{1}{(\pi (2 n+1) T)^2+\vec{p}^2+X+i\epsilon}.
\end{equation}
This paper is concerned with temperatures much larger than the masses: $T^2 \gg X$. In this case the $n=0$ \textit{zero-mode} is distinctly different from $n\neq 0$ \textit{finite modes}\te the zero-mode propagator does not depend on temperature. Note that zero-modes only appear in boson propagators.

It is well established that the loop expansion breaks down in finite temperature field theory for high enough temperatures. These temperatures are large enough to invalidate the loop counting scheme.
The worst eggs are the diagrams known as daisies~\cite{Kapusta}. They are made up of a soft momentum ($p\ll T$) inner loop, strung together with hard ($p\sim T$) self-energy insertions. Inner loops contribute $T$ and each hard self-energy contributes $\lambda T^2$, for some coupling $\lambda$. An $N$-loop daisy $d_N$ then scales as $d_N \sim T (\lambda T^2)^{N-1}$, and isn't suppressed compared to the $(N-1)$-loop daisy, to wit
\begin{equation}
  \frac{d_N}{d_{N-1}}\sim \frac{T (\lambda T^2)^{N-1}}{T (\lambda T^2)^{N-2}}\sim \lambda T^2.
\end{equation}
So perturbation theory breaks down for temperatures of order $T^2 \sim 1/\lambda$.

The well-known resolution to this problem is to perform a resummation in which bosons acquire a thermal mass. This removes all the problematic terms and replaces them with a single resummed term. Or rather, gathering up all daisies effectively resums the mass.

Now consider the implications for the finite temperature effective potential. Leading temperature corrections show up at one loop. We are interested in large temperatures, so take a high-temperature expansion for granted: $T\gg M_{\Psi}(\phi)$ for all fields $\Psi$.

In the high temperature expansion at one loop there are terms that contribute as $T^2, T, T^0, T^{-2},\mathellipsis$, which we denote as\footnote{Here and in the following we always discard terms that are independent of $\phi$.}
\begin{equation}
  V_1(\phi)= T^2 V_1^2 + T V_1^1 + V_1^0 + \mathellipsis
\end{equation}

Note that $\Vmin$ is evaluated in the same way as in the zero temperature $\hbar$-expansion. First calculate $\phim$ perturbatively and then evaluate the potential at $\phim$. The difference is that $\phim$ depends on the temperature:
\begin{equation}
  \phim=\phi_0+\hbar\ka\left(T^2 \phi_1^2+T \phi_1^1+\phi_1^0+\ldots\right)+\mathellipsis
\end{equation}
To untangle the notation a bit, consider the $T^2$ correction at $\hb$ and $\hb^2$:
\begin{equation}
 \hbar T^2 V_1^2\left|_{\phi_0}\right.+\hbar^2 T^2\left( V_2^2\left|_{\phi_0}\right.-\left[\frac{(\phi_1^1)^2}{2}+\phi_1^2 \phi_1^0 \right]\partial^2 V_0\left|_{\phi_0}\right.\right) + \ordo{}(\hbar^3).
\end{equation}
This expression is gauge invariant order-by-order in $\hb$\te{}as we have confirmed to two loops.

For a general potential the naive leading-order contributions are
\begin{align}
V=V_0&+\hbar \ka\left(T^2 V_1^2+T V_1^1+V_1^0+\ldots\right)\nonumber
\\&+\hbar^2 \ka^2\left(T^3 V_2^3+T^2 V_2^2+T V_2^1+V_2^1+\ldots\right)+\ldots,
\end{align}
where both loop counting, with $\kappa$, and naive power counting, with $\hbar$, are included.

The situation is disparate at high temperatures. The leading behaviour is set by the classical potential $ V_0 $, and the (largest) loop term is given by $ \hb T^2 V_1^2 $. The loop term can only compete with the classical potential for temperatures of order $T\sim 1/\sqrt{\hb}$, which begs for a reshuffling of the perturbative expansion\te{}analogously to the reshuffle in the Coleman-Weinberg model discussed in section~\ref{ss:CW}.

Making this power-counting manifest by rescaling $T\rightarrow T/\sqrt{\hb}$, the new expansion is
\begin{align}\label{equation:NewPowerCounting}
V\rightarrow \left(V_0 + \kappa T^2 V_1^2\right)
&+\hb^{1/2}\left(\kappa T  V_1^1 + \kappa^2 T^3  V_2^3 + \kappa^3 T^5 V_3^5 + \mathellipsis\right)
\\&+\hb\left(\kappa  V_1^0 + \kappa^2 T^2 V_2^2 + \kappa^3 T^4 V_3^4 + \mathellipsis\right)\nonumber
\\&+\hb^{3/2} \left(\kappa^2  T V_2^1 +\kappa^3  T^3 V_3^3+\mathellipsis\right)\nonumber
\\&+\mathellipsis,\nonumber
\end{align}
Higher loop terms are now as important as lower loop ones\te the harbinger of a resummation.
\subsection{Thermal resummations and power counting}
Close to the phase transition temperature the potential takes a form akin to equation~\eqref{equation:NewPowerCounting}. The minimum background energy is found by minimizing the potential order-by-order in $\hb$: $\phim=\phi_0(T)+\hbar^{1/2}\phi_{1/2}+\hbar \phi_1+\mathellipsis$ The minimization conditions are
\begin{alignat}{2}
\mathcal{O}\left(\hbar^0\right):& \hspace{1cm} \partial \left[V_0+T^2 \ V_1^2\right]\left|_{\phi_0(T)}\right.&&=0,
\\\mathcal{O}\left(\hbar^{1/2}\right):& \hspace{1cm}   \left[\phi_{1/2} \partial^2\left(V_0+T^2  V_1^2\right)+ T \partial V_1^1+ T^3 \partial V_2^3+\mathellipsis\right]\left|_{\phi_0(T)}\right.&&=0,
\\& \hspace*{3.7cm}\vdots &&\nonumber
\end{alignat}
The short-hand notation $\partial\equiv \partial_\phi$ is used extensively to avoid clutter.
Note that the leading order VeV $\phi_0(T)$ is temperature dependent, and terms starting at $\hb^{1/2}$ get contributions from all loop orders.

The energy is
\begin{align}
\Vmin &= \left(V_0+ T^2 V_1^2\right)\left|_{\phi_0(T)}\right.
+\sqrt{\hb}\left(  T V_1^1+ T^3 V_2^3 +\mathellipsis \right)\left|_{\phi_0(T)}\right.\nonumber
\\ &+\hb\left.\left( V_1^0+ T^2 V_2^2 -\frac{(\phi_{1/2})^2}{2} \partial^2\left(V_0+T^2 V_1^2\right)+\mathellipsis \right)\right|_{\phi_0(T)}+\ldots
\end{align}

This result should be gauge invariant if the power counting is consistent. As it stands it is only possible to check gauge invariance if an infinite number of diagrams are included. Serendipitously enough, it's possible to resum all terms.
It is instructive to discern why a resummation is necessary in the first place. The new counting $T \sim 1/\sqrt{\hb}$ implies that the leading-order result is an amalgamation of 1-loop terms with tree-level ones. So the inverse propagator of a particle with squared mass $x$ is enhanced,
\begin{equation}
  \Delta_x^{-1}=p^2+x+ \hb \Pi(p^2) = p^2+ x + T^2 \Pi_1^2 + \ordo{}(\sqrt{\hb}),
\end{equation}
and loop-corrections are of the same order as tree-level masses.\footnote{Something similar happens in the Coleman-Weinberg model where instead $x\sim \hb$. And both terms are again of the same order.} This is akin to the familiar hard thermal loop resummation~\cite{Kapusta}, but with minor differences.

First of all, nothing has been said about which propagator lines should be resummed. There have been arguments in the past~\cite{Arnold:1992rz,Dine:1992wr} to only resum zero-modes; because the worst divergences are removed, and diagrams are not double-counted. We take a different approach, and let gauge invariance guide the way. We have confirmed that no linear-in-$\phi$ term is generated in Abelian Higgs and the Standard Model, which is a nice consistency check~\cite{Dine:1992wr}.

There's an infinite number of terms at NLO,
\begin{equation}
  \hb^{1/2}\left(T \ka V_1^1 + T^3 \ka^2 V_2^3 +T^5 \ka^3 V_3^5 + \mathellipsis\right).
\end{equation}
These are the most divergent pieces in the daisy diagrams. And they all contribute to the resummation of the zero-mode:
\begin{equation}
  \hb^{1/2}\left(T \ka V_1^1 + T^3 \ka^2 V_2^3 +T^5 \ka^3 V_3^5 + \mathellipsis\right) \rightarrow \hb^{1/2} T \ka \overline{V_1}\phantom{}^1.
\end{equation}
Scalars and $3$D-longitudinal gauge boson have been resummed in  $\overline{V_1}\phantom{}^1$.
Recall that sub-leading terms must be evaluated at the temperature dependent minimum,
\begin{equation}
  \Vmin = \left(V_0+\ka T^2 V_1^2\right)\left|_{\phi_0(T)}\right.
  +\hb^{1/2} T \ka \overline{V_1}\phantom{}^1\left|_{\phi_0(T)}\right.+\ordo{}(\hb).
\end{equation}
As discussed in~\cite{Patel:2011th}, at one loop all gauge dependence manifests itself in the masses of the Goldstone bosons. So the gauge dependence only cancels if all Goldstone masses are identically zero. Thus in a theory with a standard loop counting the 1-loop potential would be evaluated in the temperature-independent tree-level minimum\te{}where resummed Goldstone masses are non-zero. When Goldstone masses are resummed according to the process above, they are zero in the new minimum. And hence the NLO correction to $\Vmin$ is gauge invariant. This is part and parcel of the $\hb$-expansion~\cite{Patel:2011th}.

Moving on to NNLO, some novel patterns appear. Start by considering the temperature independent 1-loop term $ \hb V_1^0$. In the Arnold-Espinosa approach~\cite{Arnold:1992rz} this term is not resummed. Yet there are good reasons to resum it. First, there are terms at $\hb$ coming from all loop orders. Second, $V_1^0$ isn't gauge invariant without a resummation. The reason is the same as for the $\hb^{1/2}$ term.

But there is another reason for resumming this term. Two-loop terms of the form $T^2 V_2^2$ have two different origins. The first is purely due to zero-modes (soft momenta) and cannot be removed by resumming. The second part comes from a mix of zero- and finite-modes. Before terms were reshuffled, the $\hb$-expansion contained terms $\sim T^2 \phi_1^0 \phi_1^2\partial^2 V_0 $. These terms are of the same form: a finite-mode contribution $\phi_1^0$, and a zero-mode contribution $\phi_1^2$. But these terms are washed away\te{}the temperature scaling pushed them higher in the expansion.

It turns out that resuming $V_1^0$ is equivalent to including the aforementioned $\hb$ terms lost by the scaling.
To wit, resuming a mass $X$ in $V_1^0$ demands a subtraction to avoid over-counting:
\begin{align}
&\overline{X}=X+\kappa T^2(\Pi_X)_1^2,
\\&V_1^0 \rightarrow  \ol{V_1}\phantom{}^0-\kappa^2T^2 (\Pi_X)_1^2 \partial_X V_1(\phi),
\end{align}
with similar subtractions at higher orders.
 To sum it up, $\smash{\ol{V_1}\phantom{}^0\big|_{\phi_0(T)}}$ is gauge invariant, and so are the remaining 2-loop terms after subtracting diagrams.

In this way all the gauge dependence of $T^2 V_2^2$ is cancelled in two steps. The resummation of $V_1^0$ removes the first chunk. And resumming at two loops ($T^2 V_2^2+\mathellipsis\rightarrow\ol{V_2}\phantom{}^2$), together with the $\hbar$ expansion, removes the last bit since Goldstone masses vanish at $\phi_0(T)$.

To be clear, we advocate that the scalar masses should always be resummed, beyond their contribution to the leading order potential. This is demanded by gauge invariance. Gauge bosons are another matter, because only $3$D-longitudinal zero-modes have a large self-energy. Hence only zero-modes of vector bosons should be resummed. We give an extended discussion about how to resum vector boson masses at higher orders in appendix~\ref{app:resum}.

%% file: tex/phases.tex
\section{Phase transitions}\label{sec:phases}
Whereas the previous section delineated how the perturbative expansion of the effective potential is reshuffled with the scaling $T\sim 1/\sqrt{\hb}$, this section applies these results to phase transitions, both first- and second-order.

To make the  discussion lucid, focus on the generic potential
\begin{equation}
  V_0(\phi)=\frac{m^2}{2}\phi^2+\frac{\lam}{4}\phi^4,
\end{equation}
with $m^2 < 0,~\lambda > 0$.

\subsection{Second-order transition}
Consider first a second-order transition. With the scaling $T\sim 1/\sqrt{\hb}$ the energy is
\begin{align}\label{eq:vm}
\Vmin&=\lb \left(V_0+ T^2 V_1^2\right)+\sqrt{\hb} T \ol{V}\phantom{}^1_1\right.
\\
&+\hb\left.\left(T^2 \ol{V}\phantom{}^2_2+\ol{V}\phantom{}^0_1-\sum_X\Pi_X \partial_X V_1^0-T^2 \frac{(\phi_{1/2}(T))^2}{2}\left(\partial^2 V_0+T^2 \partial^2 V_1^2\right) \right)+\ldots\rb\Big|_{\phi_0(T)}.\nonumber
\end{align}
The leading-order term $\left(V_0+ T^2 V_1^2\right)$ determines the temperature dependent VeV $\phi_0(T)$. Terms in $T^2 V_1^2$ are gauge invariant and are of the form $\sim e^2 \phi^2 T^2$ for some coupling $e$~\cite{Patel:2011th}. So all that changes for finite $T$  is
$m^2 \rightarrow m^2_{\text{eff}}(T)$. The transition occurs at the temperature where $m^2_{\text{eff}}(T)$ changes sign: $m^2_{\text{eff}}(\Tcross)=0.$ This is a second-order transition.

\subsection{First-order transition}
Let's for a moment forget everything about proper power-counting and just try to naively describe a first-order transition, where the minimum abruptly changes from non-zero to zero for some temperature $T_c$.
This requires a barrier to develop between the two minima. To be concrete, consider a high temperature expansion in the Abelian Higgs model. For high temperatures the potential is approximately
\begin{equation}
V(\phi)\sim-m^2 \phi^2+T^2 \phi^2 (e^2+\lam)-e^3 T \phi^3+\lam \phi^4.
\end{equation}
Following~\cite{Arnold:1992rz}, these various terms have to balance each other for a barrier to develop. The balance occurs if $\lam \phi^2\sim e^3 T \phi \sim (-m^2+T^2 e^2+\lam T^2)\equiv m^2_{\text{eff}}(T)$, or
\begin{equation}
  \phi\sim \frac{e^3}{\lam}T~\hspace{1em}\&~\hspace{1em}m^2_{\text{eff}}(T)\sim \frac{e^6}{\lam}T^2.
\end{equation}
So does this scaling always work? No. It depends on the couplings: vector bosons' thermal masses, $\sim e^2 T^2$, dominate tree-level ones if  $\lam\sim e^2$, which would break any semblance of a power-counting.\footnote{This does not mean that $\lam \sim e^2$ is in general inconsistent\te{}the scaling is fine when considering second-order transitions. However, nothing can\te in perturbation theory\te be said about first-order transitions if $\lam \sim e^2$.}

A counting like $\lam\sim e^4$\te as in the Coleman-Weinberg model\te is likewise dicey. To wit this counting implies $e \phi\sim  T$ which invalidates the high-temperature expansion. To let $\lambda$ scale as higher powers of $e$ will only worsen the problem, and lower powers than $2$ will similarly break the perturbative expansion. This leaves only one option~\cite{Arnold:1992rz},
\begin{equation}
  \lam \sim e^3:~\hspace{3em} \phi \sim T~\hspace{1em}\&~\hspace{1em}m^2_{\text{eff}}(T)\sim e^3 T^2~\hspace{1em}\&~ \hspace{1em}T\sim \frac{1}{e}.
\end{equation}
So we should really be counting powers of $e$, and be fastidious about the power-counting. In the end the first-order scaling is a hybrid between a Coleman-Weinberg-like scaling (pushes terms up in order) and the second-order scaling (drags terms down to lower orders).

There will be infinite towers of diagrams at each order in the perturbative expansion, just as for the second-order scaling. Though note that scalar masses now scale differently. For example, the resummed Goldstone mass scales as $\overline{G}\sim  m^2_{\text{eff}}(T)\sim \hb^{1/2}$. This implies that previously sub-leading Goldstone self-energy terms of order $T \hb^{1/2}$ must now be resummed. So resummed scalar masses are $\overline{X}=X+T^2 \Pi_X^2+T\Pi_X^1$, where only leading order terms are included in $\Pi_X^1$. This is quite natural since $\VLO$ includes order $T$ and $T^2$ terms; inherited by scalars through $\overline{H}=\partial^2 \VLO,~\overline{G}=\partial \VLO / \phi$.
 This does not apply to gauge bosons since their masses scale as before.

\subsubsection{First-order counting and gauge dependence}

The above discussion disregarded everything that had to do with gauge symmetry and further complications from the power counting. So it may not be surprising that a naive application of this method is gauge dependent. The effect is particularly transparent in $R_\xi$ gauges.

The $R_\xi$ effective potential is schematically
\begin{align}
V(\phi)&\sim{-m^2 \phi^2 +\lam \phi^4}\nonumber
\\&\hspace{1em}+\hb\left(T^2( \lam+e^2)-3e^3 T \phi^3 +\xi^{3/2}e^3 \phi^3 T-(\ol{G}+\xi e^2 \phi^2)^{3/2}T+\ldots\right)+\mathellipsis,
\\ \ol{G}&\sim-m^2+e^2 T^2+e^3 T \phi+\lam \phi^2,
\end{align}
where the Goldstone's zero-mode has been resummed.

The development of a barrier required for a first-order transition is driven by terms proportional to $e^3 T \phi^3$. Note that these terms vanish for $\xi=3^{2/3}$. So the gauge dependence is no paltry effect. Not only does the potential depend on $\xi$, the very nature of the phase-transition is extremely sensitive of $\xi$.

The situation is alleviated with a proper power-counting. Consider the first-order transition scaling $\lam\sim e^3,~m^2_{\text{eff}}(T)\sim e^3 T^2,T\sim \frac{1}{e}$. A new minimum develops when the quartic term competes with the mass term: $\phi \sim T$. Now, the Goldstone mass is of order $\ol{G}\sim e^3 T^2$, while the photon mass is of order $e^2 \phi^2 \sim e^2 T^2$. This means that the gauge dependent terms (to leading order) cancel, leaving
\begin{equation}
  (\ol{G}+\xi e^2 \phi^2)^{3/2}T-\xi^{3/2}e^3 \phi^3 T=\frac{3}{2}T \sqrt{\xi} e \phi \ol{G}\sim e^4 T^4.
\end{equation}
So $m^2_{\text{eff}}(T) \phi^2+\lam \phi^4\sim e^3 T^4$ while $T \ol{G} \sqrt{\xi} e \phi \sim e^4 T^4$. Gauge dependent terms are sub-leading. What's more, gauge dependent terms are evaluated at $\phi_0(T)$, and by definition vanish after a resummation: $\overline{G}\left.\right|_{\phi_0(T)}=0$. Finally, note that $(\ol{G}+\xi e^2 \phi^2)^{3/2}T$ could only be expanded because
$\ol{G}\sim \lambda \phi^2\sim e^3 T^2$. This is not true if $\lambda\sim e^2$. This is another sign that first-order transitions can only be described perturbatively if $e^2\gg \lambda$.

\subsubsection{Details of the perturbative expansion}
Due to its numerous appearances, it is felicitous to use $e$ instead of $\hb$ for counting powers. So $e$ serves bilaterally as a power and a constant\te a powerful constant indeed. Gauge bosons scale as $Z\sim e^0$, and scalars as $\ol{H},\ol{G} \sim e$. In the Standard Model for example $e\sim \sqrt{\alpha_W}\sim 0.1$.

The VeV scaling ($\phi\sim T \sim e^{-1}$) implies that the leading-order potential scales as $V_0\sim \lam \phi^4 \sim e^{-1}$. Next-to-leading order terms come from $T^2 V_2^2$ and $V_1^0$ (with scalars and powers of lambda pushed to higher orders); these terms scale as $e^0$. Cracking on, NNLO is solely due to scalar $T V_1^1$ terms.\footnote{Technically there are terms from
$T^2 V_2^2 $, but these all cancel against resummation subtractions.}
N$^{3}$LO goes as $e$ and contains terms from $T \ol{V}\phantom{}^1_2$, $T^2 \ol{V}\phantom{}^2_2$, and $ T^3 \ol{V}\phantom{}^3_3$.

The potential and VeV are
\begin{align}
  V(\phi)&=e^{-1}\VLO(\phi)+\VNLO(\phi)+e^{1/2}\VNNLO+\mathellipsis\label{eq:pot1storder},\\
  \phim &= e^{-1}\phiLO+\phiNLO+e^{1/2}\phiNNLO+\mathellipsis
\end{align}
Where $\phim$ is calculated order-by-order in $e$. Mark that a derivative with respect to $\phi$ adds a factor of $e$: $\partial \sim e$. So
\begin{equation}
\partial V(\phi)=\partial\VLO(\phi)+e\partial\VNLO(\phi)+e^{3/2}\partial\VNNLO(\phi)+\mathellipsis,
\end{equation}
implying
\begin{align}
\ordo{}(e^0):&\hspace{3em}\partial\VLO\big|_{\phiLO}=0,\\
\ordo{}(e): &\hspace{3em}\partial\VNLO\big|_{\phiLO}+\phiNLO\partial^2\VLO\big|_{\phiLO}=0,\\
&\hspace{4em}\vdots\nonumber
\end{align}
Finally, the extremum energy is
\begin{align}
  \Vmin &= e^{-1} \VLO\big|_{\phiLO} + \VNLO\big|_{\phiLO} +e^{1/2} \VNNLO\big|_{\phiLO}\nonumber \\
  &+ e\left(\VNNNLO-\frac{1}{2}\phiNLO^2\partial^2\VLO\right)\Big|_{\phiLO} + \mathellipsis\label{eq:vmin1st}
\end{align}
Schematically, a gauge boson $Z$ and its $3$D-longitudinal mode $Z_L$, and scalars $\ol{H}, \ol{G}$, contribute to the different orders of the potential as
\begin{align}
\VLO(\phi)    &\sim V_0(\phi) + \kappa T^2 \phi^2 (e^2+\lam^2) + \kappa  T (2 Z\phantom{}^{3/2} + Z_L^{3/2}),\\
\VNLO( \phi)   &\sim \kappa Z^2 + \kappa^2 e^2 T^2 Z,\\
\VNNLO( \phi)  &\sim \kappa T (\ol{G}^{3/2}+\ol{H}^{3/2}),\\
\VNNNLO(\phi)  &\sim \kappa^2 e^2 T (Z^{3/2}) + \kappa^3 e^4 T^3 (Z^{1/2}),\\
&\hspace{0.5em}\vdots\nonumber
\end{align}
where $\kappa$ denotes loops. 2-loop calculations suffice to calculate the potential to NNLO.

Now for the gauge dependence. Some features are quite clear up to NNLO. Scalar masses are determined from $\VLO$, and all terms are evaluated at $\phiLO$: Goldstone masses are zero, which removes most of the gauge dependence. Yet the expansion of the potential, $ V(\phi)=e^{-1}\VLO(\phi)+\VNLO(\phi)+e^{1/2}\VNNLO+\mathellipsis,$ is in Fermi gauge only correct when $\xi=0$. The discrepancy comes from $\xi$ dependent terms, formally starting at $e^{-1/4}$. These terms all scale with some negative $\ol{G}$ power.
When $\lambda\sim e^2$ they are removed by the $\hb$ expansion.
But when $\lambda\sim e^3$ they are cancelled by a combination of resummation subtractions and the $\hb$ expansion. Since these $\xi$ dependent terms always cancel among themselves, we've left them out of the expansion of $V(\phi)$. Although, these terms are relevant when explicitly checking gauge invariance. We also want to caution the reader that completely new divergences, compared to the second-order scaling, might appear when working with a finite $\xi$. For example, terms $~\sim \log G$ appear at intermediate steps at $\oo{e^0}$; though, these terms cancel in the end, albeit in a subtle way.\footnote{In light of these complications we suggest using Landau gauge when performing the $\hbar$ expansion.}

The next check would come at N$^{3}$LO and requires knowing the effective potential to three loops in a general gauge. It was only recently that the 2-loop effective potential was found for a general gauge~\cite{Martin:2018emo}. So we'll cross that bridge when we come to it. In section~\ref{sec:examples} we calculate observables in the Abelian Higgs model and in the Standard Model with the first-order scaling outlined above, and point out possible complications and pitfalls.

\subsection{Perturbative determination of $T_c$}\label{ss:tc}
A phase transition between two phases A and B occurs when

\begin{align}
&V(\phi,T)|_{\phi_A,T_c}-V(\phi,T)|_{\phi_B,T_c}=0,
\\&\partial V(\phi,T)\left.\right|_{\phi_A,T_c}=0,
\\&\partial V(\phi,T)\left.\right|_{\phi_B,T_c}=0.
\end{align}
Or $V_A=V_B$ for short.

Since both $V_A$ and $V_B$ are gauge invariant by themselves, $T_c$ is guaranteed to be gauge invariant. There are two schools of thought on how to find the critical temperature. First, draw the phase-diagram for $V_A$ and $V_B$, and change $T$ until the two energies match up. This is the method proposed in~\cite{Patel:2011th}, and it has the advantage that higher order corrections are easy to include. Though this method is gauge invariant, perturbative orders are again muddled.

An alternative is to find $T_c$ order-by-order in $\hb$,
\begin{equation}
  T_c=T_0+\hb T_1+\hb^2 T_2+\mathellipsis,
\end{equation}
as investigated by Laine~\cite{Laine:1994zq}. But he noticed that an $\hb$ expansion for $T_c$  breaks down at $\hb^2$, and so the idea has long been dismissed. Yet this breakdown does not occur for all power-counting schemes.

Consider first the second-order scaling,
\begin{align}
\Vmin&=\lb \left(V_0+ T^2 V_1^2\right)+\sqrt{\hb} T \ol{V}\phantom{}^1_1\right.\nonumber
\\&+\hb\left.\left(T^2 \ol{V}\phantom{}^2_2+\ol{V}\phantom{}^0_1-\sum_X\Pi_X \partial_X V_1^0-T^2 \frac{(\phi_{1/2}(T))^2}{2}\partial^2 V_0(T)+ \right)+\ldots\rb\Big|_{\phi_0(T)}.
\end{align}
The leading order energy vanishes in the symmetric phase: $\VLO^A=0$. For the broken phase the leading-order energy is proportional to the Higgs' temperature-dependent mass: $\VLO^B\propto H(T)^2$. So enforcing $V(\phi,T)|_{\phi_A,T_c}-V(\phi,T)|_{\phi_B,T_c}=0$ at leading order gives $H(T)|_{T_0}=0$\te causing problems at two loops. The 2-loop potential contains terms of the form $T^2 \log H(T)$; since we'll expand around $T_0$ in the $\hb$ expansion these terms diverge and do not cancel between phase
$A$ and $B$. The expansion seems useless. In our mind this cements that the scaling $\lambda\sim e^2$ cannot describe a first-order phase transition, and that the critical temperature is then simply determined from the leading order potential\te{}with higher order corrections incalculable in perturbation theory.

Yet a first-order transition is different. The effective Higgs mass $H(T)$ is finite both in the symmetric and broken phase, and no divergences can (naively) appear. Our explicit calculations, reviewed in section~\ref{sec:results}, show that first-order transitions are free of these subtleties.
The expansion of the critical temperature is of the form $T_c=e^{-1}\TLO+e \TNLO+e^{3/2}\TNNLO+\ordo{}(e^2)$. Derivatives with respect to $T$ scale as $e^0$ when acting on $\overline{G}$, $\overline{H}$ or $\VLO$, and as $e$ when acting on anything else.

Denote the potential difference as $\Delta V(\phi)\define V(\phi)-V(0)$, whose expansion is
\begin{align}
  0=\Delta V(\phim)\big|_{T_c}&=e^{-1}\Delta\VLO\big|_{\phiLO,\TLO}\nonumber
  \\& + e^{0}\left(\Delta\VNLO+ \TNLO\partial_T\Delta\VLO\right)\Big|_{\phiLO,\TLO}\nonumber
  \\& +e^{1/2}\left(\Delta\VNNLO+\TNNLO\partial_T\Delta\VLO\right)\Big|_{\phiLO,\TLO}
  \\& ~\,\vdots\nonumber
\end{align}
Note that $\partial_T\Delta\VLO$ scales as $e^{-1}$, which is why $\TNLO$ scales as $e$.\footnote{The $T$-derivative breaks apart the careful balance in $m_{\text{eff}}^2(T)$, enhancing the scaling.} The additional suppression ($\TNLO/\TLO\sim e^2$) explains why corrections to $\TLO$ tend to be rather small, as seen in section~\ref{sec:results}.
With $T_c$ found it is possible to calculate various observables at the phase transition. For example, the barrier height is
\begin{align}
  \Vbarr &= e^{-1} \Delta\VLO\big|_{\phibarr, \TLO} + e^{0}\left(\Delta\VNLO+\TNLO\partial_T\Delta\VLO\right)\Big|_{\phibarr, \TLO}\nonumber
  \\&+e^{1/2} \left(\Delta\VNNLO+\TNNLO\partial_T\Delta\VLO\right)\Big|_{\phibarr, \TLO} +\mathellipsis
\end{align}
Where $\phibarr$ is the location of the leading-order maximum defined by
\begin{equation}
\partial\VLO\big|_{\phibarr}=0.
\end{equation}

Note that our calculation entails first expressing everything at $\phiLO(T)$, and then expanding $T=\TLO+\TNLO+\ldots$
For a function $F(\phi,T)$ the expansion around $T_0$ then contributes two types of terms: explicit and implicit derivatives with respect to $T$. To wit consider expanding $F(\phi,T)$ first around $\phi=\phiLO(T)+\phiNLO(T)+\ldots$, and then around $ T=\TLO+\TNLO+\mathellipsis$,
\begin{align}
F(\phi,T)=\left\{F+\TNLO \partial_T F+\partial_T \phiLO \partial_\phi F+\phiNLO \partial_\phi F\ldots\right\}\Big|_{\phiLO,\TLO}.
\end{align}
Temperature derivatives of $\phiLO(T)$ can be rewritten as $\partial_T \phiLO=-\partial_T \partial_\phi \VLO/\partial_\phi^2\VLO$ and similarly for higher orders. So everything boils down to an \textit{effective} $\phiNLO$: $\phiNLO\rightarrow \phiNLO-\TNLO \partial_T \partial_\phi \VLO / \partial_\phi^2\VLO$. The new $\phiNLO$ automatically takes care of all implicit derivatives. So when expanding around $T_c$ we'll always use the temperature corrected $\phiNLO$.
\subsection{Thermodynamical Observables}
Finding the critical temperature is all well and good, but there are a myriad of other observables. For example, the sphaleron transition rate is approximately controlled by $\frac{v_c}{T_c}$~\cite{Patel:2011th}, where $v_c$ is taken to be the minimum at the critical temperature. Larger values of this ratio indicate that sphaleron proccesses are suppressed enough after the phase transition to not erase any generated matter-antimatter asymmetry. Such phase transitions are \emph{strongly first-order} if $v_c/T_c \geq 1$. If this measure is to be physically meaningful, it has to be gauge independent. But the minimum, numerically found or perturbatively expanded, is gauge dependent and IR-divergent, and so a rather bad observable. Though there is a related observable, the scalar square vacuum expectation value $\left\langle\, \left|\Phi\right|^2\right\rangle$.
The scalar square expectation value is gauge invariant in the minimum, and is given by
\begin{align}\label{eq:squareVev}
\left\langle \,\left|\Phi\right|^2\right\rangle=2\frac{\partial}{\partial m^2}V(\phi,T),
\end{align}
where $m^2$ is the mass in the scalar potential~\cite{Laine:2017hdk}.\footnote{Note that the numerical coefficient in equation~\eqref{eq:squareVev} is convention dependent. It is chosen such that the leading order contribution is $\phiLO^2$.}
Although this quantity is scale dependent, the scale can be nailed down when comparing against lattice~\cite{Farakos:1994kj}. So we take
$\left\langle\, \left|\Phi\right|^2\right\rangle / T_c$ as a proxy for the sphaleron transition rate.

Let's see how to find $W\equiv 2\frac{\partial}{\partial m^2}V(\phi,T)$ order-by-order.  Take the potential as
\begin{align}
V(\phi,T)=e^{-1}\VLO(\phi,T)+e^0 \VNLO(\phi,T)+e^{1/2}\VNNLO(\phi,T)+\ldots,
\end{align}
where all resummations are included as in the previous section. And equivalently
\begin{align}
&W=e^{-2}\WLO(\phi,T)+e^{-1} \WNLO(\phi,T)+e^{-1/2} \WNNLO(\phi,T)+\ldots,
\\& W_i=2\frac{\partial}{\partial m^2}V_i.
\end{align}
The expansion is
\begin{align}
W_{\text{min}}=\left. e^{-2}\WLO\right|_{\phiLO} &+\left.e^{-1}\left[\WNLO+\phiNLO \partial \WLO\right]\right|_{\phiLO}\nonumber
\\&+\left.e^{-1/2}\left[\WNNLO+\phiNNLO \partial \WLO\right]\right|_{\phiLO}+\mathellipsis
\end{align}
We straight off the bat see a different story than for $\Vmin$: $\phiNLO$ contributes already at NLO\te{}this didn't happen for $\Vmin$ because $\partial \VLO\left.\right|_{\phiLO}=0$.

We have verified that $W$ is gauge invariant and finite up to order NNLO.

The phase transition strength is then
\begin{align}
\frac{\sqrt{W}}{T_c}&=\frac{\sqrt{\WLO}}{\TLO}\left[1+e \frac{\WNLO+\phiNLO \partial \WLO}{2 \WLO}+e^{3/2} \frac{\WNNLO+\phiNNLO \partial \WLO}{2 \WLO}+\ldots\right].
\end{align}
With effective VeV $\phiNLO\rightarrow \phiNLO-\TNLO \partial_T \partial_\phi \VLO / \partial_\phi^2\VLO$ and similarly for $\phiNNLO$.

On the other hand of the spectrum there's the \textit{latent heat}, denoted $L(\phi,T)$, and defined as $L \define T \partial_T \Delta V$~\cite{Laine:1994zq}:
\begin{align}
L(\phi,T)&=e^{-2}\LLO(\phi,T)+e^{-1}\LNLO(\phi,T)+e^{-1/2}\LNNLO(\phi,T)+\mathellipsis,
\\ L_{\text{min}}&=\left.e^{-2}\LLO\right|_{\phiLO}
+\left.e^{-1}\left[\LNLO+\phiNLO \partial\LLO\right]\right|_{\phiLO}\nonumber
\\&\hspace{5.85em}+\left.e^{-1/2}\left[\LNNLO+\phiNNLO \partial\LLO \right]\right|_{\phiLO}+\mathellipsis
\end{align}

\subsection{Summary of procedure}
With the considerations of the previous subsections out of the way, we can now summarize the recipe for calculating the vacuum energy at finite temperature. The first thing to consider is the size of the four-point coupling $\lambda$ compared to the coupling of other bosons (we use a generic gauge coupling $e$ to facilitate discussions).

If $\lambda \sim e^2$ a second-order transition takes place at leading order. The perturbative expansion of $T_c$ seems to break down for higher orders. The critical temperature $\Tcross$ can be readily found from the leading-order potential. In our notation it can be written
\begin{equation}
  \left(\Tcross\right)^2=-\frac{\partial^2V_0}{\partial^2 V_1^2}\Big|_{\phi=0}.
\end{equation}

If $\lambda \sim e^3$, then a first-order phase transition is possible. If this is the case, finding the critical temperature requires performing a perturbative expansion\te{}there are a number of steps. In the high-temperature expansion, a bosonic mode $X$ contributes to the 1-loop effective potential as
\begin{equation}
 \frac{1}{12}T^2 X-T\frac{1}{4 \pi} X^{3/2}-\frac{X^2}{64 \pi^2}\log\left[ \frac{e^{2\gamma_E}Q^2}{16 \pi^2 T^2} \right]+\mathellipsis,
\end{equation}
while a fermionic mode contributes with a similar $T^2$ and $T^0$ term but no $T$ term.
\begin{enumerate}
  \item The leading order potential is
    \begin{equation}
      \VLO(\phi) = V_0(\phi)+\frac{1}{12}T^2 \sum_X X-\frac{T}{4 \pi}\left(\sum_{X'} X'+\sum_{\ol{X}}+\ol{X}\right),
    \end{equation}
    where the sum over $X$ ranges over all particles, the sum over $X'$ over bosons whose masses scale as $e^0$ (i.e. not scalars) but that are not resummed, and the sum over $\ol{X}$ are for $e^0$-scaling masses that \emph{are} resummed ($3$D-longitudinal modes).
  \item To find higher orders (NLO, NNLO, \ldots) use the following scaling rules. Perform the high-temperature expansion for each loop order. The scalar squared masses count as a factor of $e$, other squared masses as $e^0$. Each factor of $T, \phi$ (outside masses) count as $e^{-1}$.
  \item Excluding the $T^2V_1^2$ term, all scalar masses should be resummed. The new masses are found from the leading-order potential, e.g. $\ol{H}=\partial^2 \VLO, \smash{\ol{G}=\partial \VLO/\phi}$. For gauge bosons, only their $3$D-longitudinal parts are resummed, and only the zero-modes. Gauge boson self-energies must be explicitly calculated.
  \item Perform resummation subtractions. For example, replace the squared mass $X$ by its resummed version $\ol{X}=X+\kappa \Pi_X$ in the integral function $h(x)$, and then subtract off the generated extra terms,
     \begin{equation}
       h(X)=h(\ol{X})-\kappa \Pi_X h'(X)-\kappa^2 \frac{1}{2}\Pi_X^2 h''(X)+\mathellipsis
     \end{equation}
     However, this requires singling out the terms that should not be resummed, and has to be done while resumming and in a specific order. In practice it is simpler to insert the resummed masses first and then also subtract off all double countings,\footnote{This is equivalent to using a thermal counter-term~\cite{Parwani:1991gq}.}
          \begin{equation}
            h(X)=h(\ol{X})-\kappa \Pi_X h'(\ol{X})-\kappa^2 \frac{1}{2}\Pi_X^2 h''(\ol{X})+\mathellipsis
          \end{equation}

  \item Find $\phim$ in the perturbative expansion by solving $\partial V = 0$ order by order in $e$. Each $\phi$ derivative scales as $e$.
  \item Evaluate $\Vmin$ perturbatively by expanding $\phim$ in the expression for $V$.
  \item Find $T_c$ by solving $\Vmin=V(0)$ either by varying $T$ continuously or by performing a perturbative expansion (our recommendation). When expanding $T_c$ use the effective VeV to include implicit $T$ derivatives: $\phiNLO\rightarrow \phiNLO-\TNLO \partial_T \partial_\phi \VLO / \partial_\phi^2\VLO$. And $T$ derivatives scale as $e^0$ when acting on $m^2_{\text{eff}}$, and as $e$ when acting on anything else.
  \item Other observables are found by similarly expanding around $T_c$ and $\phim$.
\end{enumerate}
\newpage

%% file: tex/examples.tex
\section{Examples}\label{sec:examples}
This section shows explicit calculations; we'll use Abelian Higgs and the Standard Model as guinea pigs.

\subsection{The effective potential}\label{ss:effpot}
Our aim is to be accurate and have confidence in our accuracy. To test gauge dependence, and by extension our power-counting, we need to calculate the effective potential to 2-loop order in a generic gauge. Thankfully, such calculations are tractable since the 2-loop effective potential for a generic model, and gauge, is known at zero temperature~\cite{Martin:2018emo}. We extend these results to finite temperature and calculate some observables to NNLO accuracy.

The calculation is structured so that everything depends solely on a number of master integrals. Unrenormalized master integrals are distinguished from renormalized ones by a boldface. For example, the unrenormalized 1-loop thermal integral, depending on a bosonic mass $X$, is $\fB{X}$\te $T$ dependence is left implicit. High-temperature expansions of all master integrals are given in appendix~\ref{app:integrals}.

The 1-loop effective potential is a mere functional determinant. And so contributions from different fields separate. Letting $X$ represent bosons, and $Y$ fermions, the generic result is
\begin{equation}
  V_1=\sum_X \fB{X}-2\sum_Y \fF{Y}.
\end{equation}
Two-component fermions bring an additional minus sign and a factor of two. Fermions and bosons have different master integrals. Denote the fermion one by $\fF{x}$.

There are sundry diagrams at two loops\te{}see figure 3.1 in~\cite{Martin:2018emo} for all possibilities. For example, the contribution of two scalars and a mixed scalar-vector is
\begin{equation}
  \frac{1}{2}\lam^{A\mathbf{jk}}g^{A}_{\mathbf{jk}}a_\mathbf{j}^{\epsilon}a_\mathbf{k}^{\epsilon'}c_{A}^{\epsilon''}\mathbf{f}_{SSG}(\mathbf{j}_{\epsilon},\mathbf{k}_{\epsilon'},A_{\epsilon''})+\mathellipsis,
\end{equation}
where $\lambda, g, a, c$ are combinations of couplings and masses, and $\mathbf{f}_{SSG}$ is an \textit{integral function} given in terms of master integrals and masses.

At two loops there are are also contributions from counterterm insertions in the 1-loop potential. We opt to perform these insertions explicitly, rather than using the renormalized integral functions of~\cite{Martin:2018emo}.

We review a few integral functions in appendix~\ref{app:intfuncs}; these deviate slightly from~\cite{Martin:2018emo} with the inclusion of new temperature pieces.

We perform several cross-checks to ensure the validity of our results. Both in Abelian Higgs and the Standard Model. We have tested gauge invariance in the standard loop counting and in the modified ones; renormalization group invariance; and the removal of the "dangerous" 2-loop $T^3 \phi$ term~\cite{Dine:1992wr}.

\subsection{Abelian Higgs}
\subsubsection{Model definition}
We use the same conventions as~\cite{Martin:2018emo} to make comparisons easy. The Abelian Higgs model is defined by the Lagrangian
\begin{equation}
\lag=-\frac{1}{4}F_{\mu \nu}F^{\mu \nu}-\left(D^{\mu}\Phi\right)^{\dagger}D_{\mu}\Phi - \left(m^2 \left|\Phi\right|^2+\lam \left|\Phi\right|^4\right)+\lag_{\mathrm{g.f.}} +\lag_{\mathrm{ghost}},
\end{equation}
where $F_{\mu\nu}=\partial_\mu Z_\nu-\partial_\nu Z_\mu$ is the field-strength of the \uone{} gauge field $ Z^{\mu} $, and $\Phi =\smash{\frac{1}{\sqrt{2}}\left(\phi_{1}+i\phi_2\right)}$ is a complex scalar field charged under this \uone{}, with covariant derivative $D_\mu\Phi = (\partial_\mu-i e Z_\mu)\Phi$.
The parameters in the scalar potential satisfy $m^2<0, \lambda > 0 $, so that there is spontaneous symmetry breaking at tree-level. Focus on Fermi gauge. The gauge-fixing and ghost terms are
\begin{align}
\lag_{\mathrm{g.f.}}&=-\frac{1}{2 \xi} \left(\partial_{\mu} Z^{\mu}\right)^{2},\\
\lag_{\mathrm{ghost}}&=-\overline{\eta}\partial^{\mu}\partial_{\mu} \eta.
\end{align}
Expand the scalar field $\Phi$ around its VeV $\phi$ as
\begin{equation}
  \Phi(x) = \frac{1}{\sqrt{2}}\left(\phi+H(x)+i G(x)\right),
\end{equation}
where $H$ and $G$ are real scalar fields. This gives the tree-level potential $V_0(\phi)=(1/2)m^2 \phi^2+(1/4)\lam \phi^4$. Using the convenient notation that the squared mass of a particle is denoted with the name of that field, the squared masses\te as functions of $\phi$\te are
\begin{align}
  H &= m^2+3\lam \phi^2,\\
  G &= m^2+ \lam \phi^2,\\
  Z_{\pm}&=\frac{1}{2}\left(G\pm\sqrt{G(G-4 \xi Z)}\right),\\
  Z &= e^2 \phi^2.
\end{align}
Ghosts are massless in Fermi gauges.

The Goldstone field, $G$, mixes with the gauge boson, $Z$; propagators of $G$, $Z$, and mixed $G$-$Z$ are expressed in terms of the gauge-dependent masses $Z_{\pm}$, which fulfill
\begin{equation}
  Z_{+}(\phi)+Z_{-}(\phi)=G.
\end{equation}
In addition, $Z_{\pm}(\phi_0)=0$, while $Z_{+}(0)=G|_{\phi=0},~Z_{-}(0)=0$. These relations ensure that the effective potential is gauge invariant when expanded around the tree-level extrema.

\subsubsection{The perturbative expansion}
The 1-loop potential is found by summing over all fields,
\begin{equation}
  \mathbf{V_1}(\phi)=\fB{H}+\fB{Z_{+}}+\fB{Z_{-}}+(3-2\epsilon)\fB{Z}.
\end{equation}
Contributions from ghosts are ignored because ghost masses are $\phi$ independent. After renormalization the result is
\begin{equation}
  V_1(\phi)=f(H)+f(Z_{+})+f(Z_{-})+3f(Z)-2f_{-1}(Z),
\end{equation}
where un-bolded functions are finite. They are given in appendix~\ref{app:integrals}.

The most important terms in $V_1$ are
\begin{align}
  V_1(\phi)&=\frac{T^2}{12}\left(H + 3 Z + Z_{+} + Z_{-}\right) +\mathellipsis\nonumber\\
          &=\frac{T^2}{12}\left(3 e^2+4\lam \right)\phi^2+\mathellipsis,
\end{align}
where $\phi$ independent terms are ignored. As emphasized in~\cite{Patel:2011th}, the $T^2$ term in $V_1$ is gauge invariant.

If the scalar coupling $\lambda$ scales as $\lam\sim e^2$, a second-order phase transition takes place. The leading-order potential for this scaling is
\begin{equation}
  \VLO(\phi)\big|_{\text{2nd-order}}=\frac{1}{2}\left(m^2+\frac{T^2}{12}(3 e^2 + 4 \lam)\right)\phi^2+\frac{1}{4}\lam \phi^4.
\end{equation}
The critical temperature is reached when the $\phi^2$ term vanishes, giving
\begin{equation}
 \Tcross = \sqrt{\frac{- 12 m^2}{3 e^2+4 \lambda}}.
\end{equation}

On the other hand if $\lam\sim e^3$, a first-order phase transition can take place. Finding the leading-order potential (and higher orders) is then more involved; additional terms from $V_1$ become important:
\begin{equation}
-\frac{T}{12 \pi}\left(H^{3/2}+Z_{+}^{3/2}+Z_{-}^{3/2}+3 Z^{3/2}\right).
\end{equation}
The leading-order potential for this scaling is
 \begin{align}
 \VLO(\phi)&=\frac{1}{2}\left(m^2+\frac{T^2}{12}(3 e^2 + 4 \lam)\right)\phi^2-\frac{T}{12 \pi}\left(2 Z^{3/2}+Z_L ^{3/2}\right)+\frac{1}{4}\lam \phi^4,
 \\Z_L &=Z+\frac{1}{3}e^2 T^2.
 \end{align}
Masses are of order $Z, Z_L\sim e^0, H\sim e, Z_{\pm}\sim e$ close to the minimum because $\phi\sim e^{-1}$. A resummation is needed. Both scalar masses are resummed in one sweep via $H\rightarrow \partial^2 \VLO, G\rightarrow\frac{1}{\phi}\partial \VLO$.
The resummed potential describes a first-order phase transition and the critical temperature is known analytically to leading order.

Next, sub-leading corrections. Both $V_1^0$ and $V_2^2$ contribute at NLO. Scalar and longitudinal vector masses are resummed as in appendix~\ref{app:resum}. Counterterm insertions also contribute:	 If a mass $X$  is renormalized in dimensional regularisation by $Z_X$, the finite counter-term contribution is
	\begin{align}
	Z_X&=\kappa \frac{1}{\epsilon} Z_X^1+\mathellipsis,\\
	\fB{ Z_X X}&=f(X)+\kappa Z_X^1 f'_\epsilon(x)+\mathellipsis
	\end{align}

The critical temperature is found in powers of e, $T_c=e^{-1}\TLO+\e \TNLO+e^{3/2}\TNNLO+\mathellipsis$, as in subsection~\ref{ss:tc},
\begin{equation}
  \TNLO = -\frac{\Delta \VNLO}{\partial_T \Delta \VLO}\Big|_{\phiLO, \TLO}.
\end{equation}
Only scalar $T V_1^1$ terms contribute at NNLO, and these terms are gauge dependent. But to find $\Vmin$, $\VNNLO$  is evaluated at $\phiLO$\te where the Goldstone mass vanishes. So the result is gauge invariant. The corresponding contribution to the critical temperature is
\begin{equation}
  \TNNLO = -\frac{\Delta \VNNLO}{\partial_T \Delta \VLO}\Big|_{\phiLO, \TLO}.
\end{equation}
Higher orders require 3-loop calculations.
\subsection{Standard Model}\label{ss:SM}
\subsubsection{Model definition}
Free parameters are $m^2$ and  $\lam$ in the scalar sector; $g_3,g,g'$  $\left(\text{SU}_c(3),\text{SU}_L(2),\text{U}_Y(1)\right)$ in the gauge sector; the top-quark Yukawa coupling is $y_t$ in the Yukawa sector. We use Fermi gauge with the gauge-fixing parameters $\xi_\gamma, \xi_Z,\xi_W$; the effective potential is independent of the gauge-parameter of the gluon, $\xi_c$, to two loops. In our numerical calculations $\xi_\gamma, \xi_Z,\xi_W\define \xi$.

The field content for $n_G$ generations is
\begin{itemize}
  \item[Vectors:] $A,~ Z,~ W_R,~ W_I,~ g_8$,
  \item[Scalars:] $H,~ G_0,~ G_I,~ G_R$,
  \item[Ghosts:] $\eta_A,~ \eta_Z,~ \eta_{W_R},~\eta_{W_I}$,
  \item[2-comp. Weyl fermions:] $t,~ \ol{t},~ b,~ \ol{b},~ \tau,~ \ol{\tau},~ \nu_\tau+(n_G-1)\times (u,~ \ol{u},~ d,~ \ol{d},~ e,~ \ol{e},~ \nu_e)$,
\end{itemize}
where $g_8$ denotes the color octet gluons.\footnote{There are technically ghosts $\eta_{g_8}$ corresponding to the gluons, but they do not contribute any $\phi$-dependence at this order in perturbation theory.} All bosons are real with the index $R$ denoting the real part and $I$ the imaginary part of the corresponding complex field. The Goldstone $G_0$ corresponds to the longitudinal mode of $Z$; $G_I$ to that of $W_R$; and $G_R$ to that of $W_I$. Though the real and imaginary part of a field have the same squared mass, the
 $R$ and $I$ labels are handy for calculations~\cite{Martin:2018emo}.

Squared masses are
\begin{align}
   H&=m^2+3\lam \phi^2,
\\ G&=m^2+ \lam \phi^2,
\\ Z&=(g^2+g'^2)\phi^2/4,
\\ Z_{\pm}&=\frac{1}{2}\left(G\pm\sqrt{G(G-4 \xi_Z Z)}\right),
\\ W&=g^2\phi^2/4,
\\ W_{\pm}&=\frac{1}{2}\left(G\pm\sqrt{G(G-4 \xi_W W)}\right),
\\ t&= y_t^2 \phi^2/2.
\end{align}
Ghosts are massless and only the top-quark mass is significant among the fermions.

The Goldstone-gauge mixing masses $Z_{\pm}, W_{\pm}$ satisfy properties analogous to those of the mixing masses in Abelian Higgs.

\subsection{The perturbative expansion}
The barrier height is given by the coefficient of the $\phi^3$ term. In Abelian Higgs this corresponded to the gauge coupling $e$. In the Standard Model, linear $T$ terms come from $W$ and $Z$ bosons: $W^{3/2}=(g/2)^3 \phi^3, Z^{3/2}=(\sqrt{g^2+g'^2}/2)^3\phi^3$. Both coefficients are of similar numerical size; we can count the gauge couplings as $g, g'\sim e$.

The renormalized 1-loop potential is
\begin{align}
  V_1(\phi)&=f(H)-12 f_F(t) + 2\left[3 f(W) - 2f_{-1}(W)\right] + 2\left[f(W_{+}) + f(W_{-})\right]\nonumber
  \\&+ 3 f(Z) - 2 f_{-1}(Z) + f(Z_{+}) + f(Z_{-}).
\end{align}
With leading terms
\begin{align}
T^2V_1^2&=\frac{T^2}{24}\left(H -12 \times\left(-\frac{1}{2}\right)t + 6 W +2 G + 3 Z + G\right)\nonumber\\
        &=\frac{T^2}{32}(8 \lam + 4 y_t^2 + 3 g^2 + g'^2)\phi^2.
\end{align}
For a second-order transition $\lambda\sim e^2$, the leading order potential is
\begin{equation}
  \VLO(\phi)\big|_{\text{2nd-order}} = \frac{1}{2}\left(m^2+\frac{T^2}{16}\left(8 \lam + 4 y_t^2 + 3 g^2 + g'^2\right)\right)\phi^2+\frac{1}{4}\lam \phi^4.
\end{equation}
With critical temperature
\begin{equation}\label{eq:smt2nd}
\Tcross = \sqrt{\frac{-16 m^2}{8 \lam + 4 y_t^2 + 3 g^2 + g'^2 }}.
\end{equation}
For a first order scaling ($\lam\sim e^3$) the leading order potential is
\begin{align}
  \VLO(\phi) &= \frac{1}{2}\left(m^2+\frac{T^2}{16}\left(8 \lam + 4 y_t^2 + 3 g^2 + g'^2\right)\right)\phi^2\nonumber
  \\&-\frac{T}{12 \pi}\left(2\left[2W^{3/2}+W_L ^{3/2}\right]+2Z^{3/2}+Z_L ^{3/2}+A_L^{3/2}\right)+\frac{1}{4}\lam \phi^4.
\end{align}
The potential can not be minimized analytically, but the minimum is found numerically in a breeze. Higher-order corrections to the critical temperature are found as in Abelian Higgs.

Though there is a new complication beyond 1-loop: $3$D-longitudinal modes of $Z$ and $A$ mix. See appendix~\ref{app:longitudinal} for the details.

%% file: tex/results.tex
\section{Numerical results in the Standard Model}\label{sec:results}
In this section we report on the numerical results from applying our method to the Standard Model, and we compare it with the traditional method of numerically minimizing the potential. The calculation is performed using the high-temperaure expansions given in appendix~\ref{app:integrals}, and the organizational framework of~\cite{Martin:2018emo}; we use Fermi gauge throughout.

We take the input parameters to be~\cite{Tanabashi:2018oca}
\begin{align}
  Q_0        &= M_Z = 91.2 \gev,\\
  y_t(Q_0)    &= 0.995,\\
  g_3(Q_0)   &= 1.28,\\
  g(Q_0)     &= 0.654,\\
  g'(Q_0)    &= 0.350,\\
  G_F    &= 1.17\times 10^{-5} \gev^{\,-2}.
\end{align}
These parameters correspond to bare masses and VeV
\begin{align}
  M_Z &= 91.2 \gev,\\
  M_W &= 80.4 \gev,\\
  M_t &= 173 \gev,\\
  v &= 246 \gev.
\end{align}
When we vary the Higgs mass while keeping the VeV $v$ fixed, we need to vary the potential parameters $\lambda,~m^2$ in tandem, since
\begin{align}
  m^2 &= - \frac{1}{2}M_H^2,\\
  \lam &= \frac{M_H^2}{2 v}.
\end{align}
As a reference we consider a benchmark point,
\begin{align}
  \lam (Q_0) &= 0.0167,\\
  m^2(Q_0)   &= -(31.8\gev)^2,
\end{align}
which gives the Higgs mass $M_H=45.0\gev$.
\subsection{Traditional method}
The traditional method finds the critical temperature $T_c$ by first calculating the effective potential to a given loop order, minimizing the potential numerically, and changing the temperature till the symmetric and broken energies coincide.

We performed this calculation for the Standard Model, using the parameters laid out above. For the 1-loop potential we elected to perform the high-temperature expansion to order $\ordo{}(x^2 T^0)$ instead of using the numerical integrals directly, noting that the expansions are accurate in the temperature range we are considering. 2-loop sunset diagrams are truncated at $\ordo{}(x T^2)$. Scalar and $3$D-longitudinal vector masses are all resummed with the leading-order-$T^2$ self-energy; the resummation targets zero-modes and the scalar $\ordo{}(x^2 T^0)$ contributions. We subtract the relevant diagrams from the 2-loop potential to prevent double-counting.

\begin{figure}
	\centering
	\captionsetup[subfigure]{oneside,margin={0.5cm,0cm}}
	\subfloat[][]{\includegraphics[width=.47\textwidth]{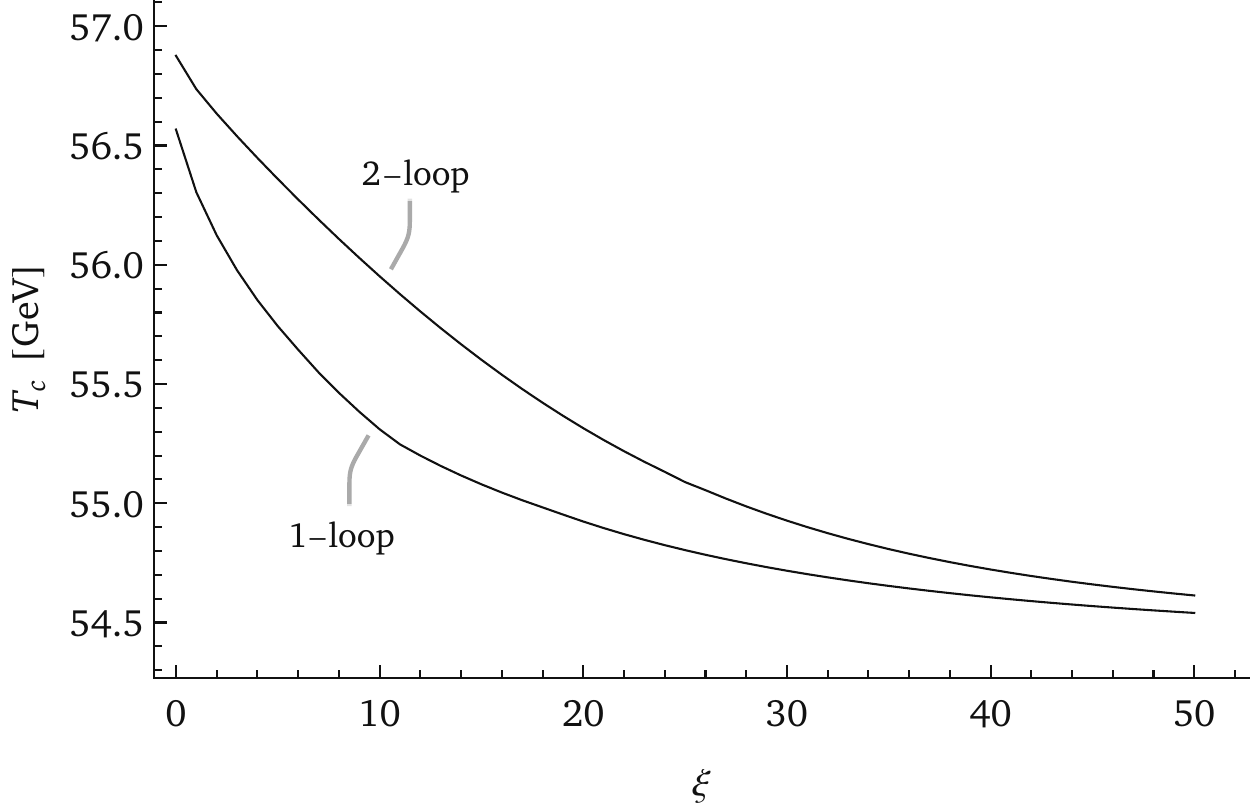}\label{sfig:TradVsXi_tc}}\quad
	\subfloat[][]{\includegraphics[width=.47\textwidth]{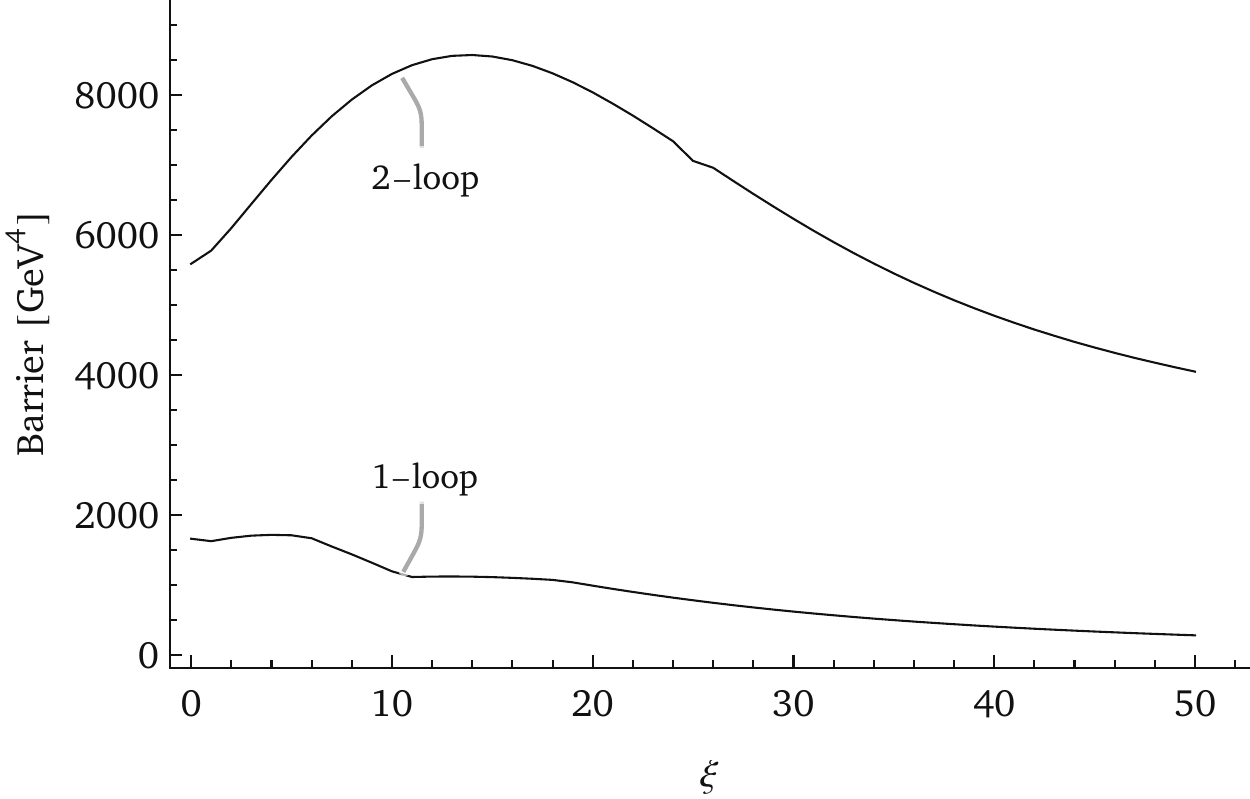}\label{sfig:TradVsXi_barr}}\\
	\subfloat[][]{\includegraphics[width=.47\textwidth]{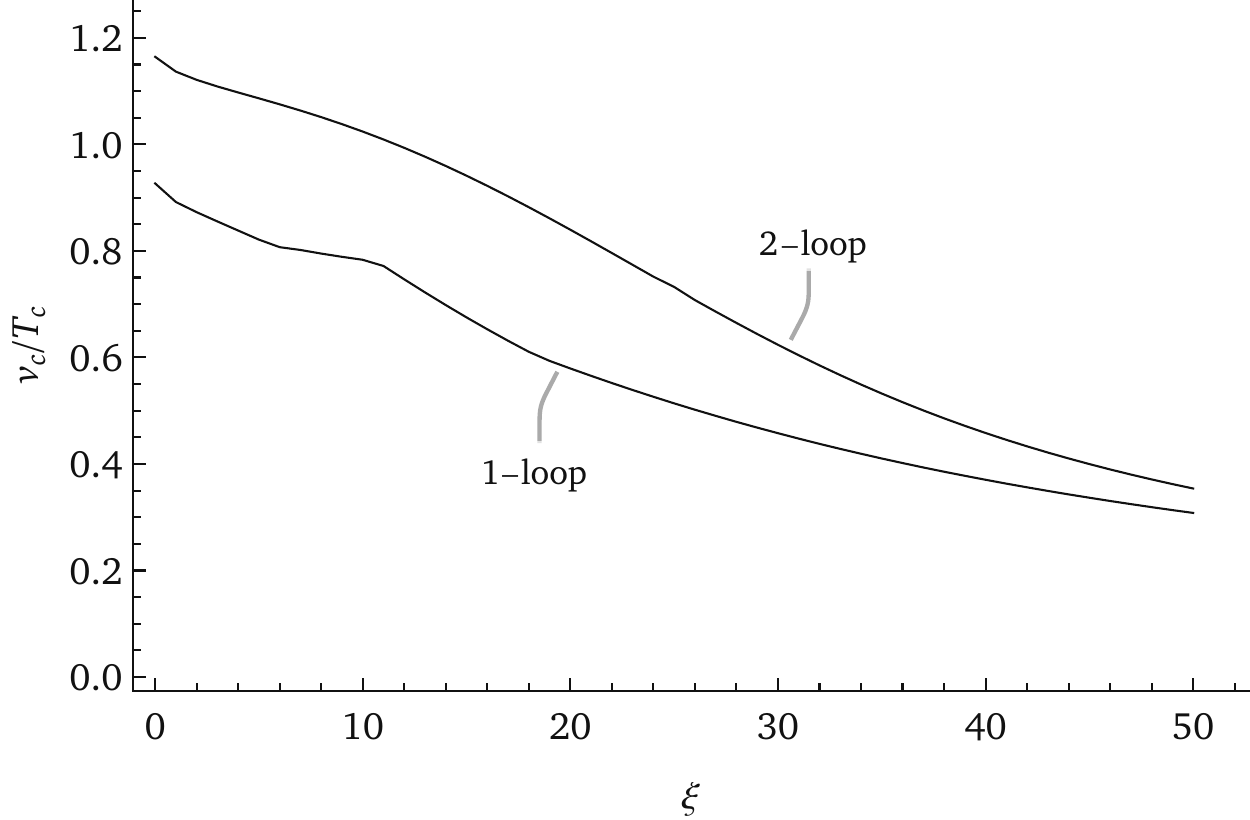}\label{sfig:TradVsXi_str}}\quad
	\subfloat[][]{\includegraphics[width=.47\textwidth]{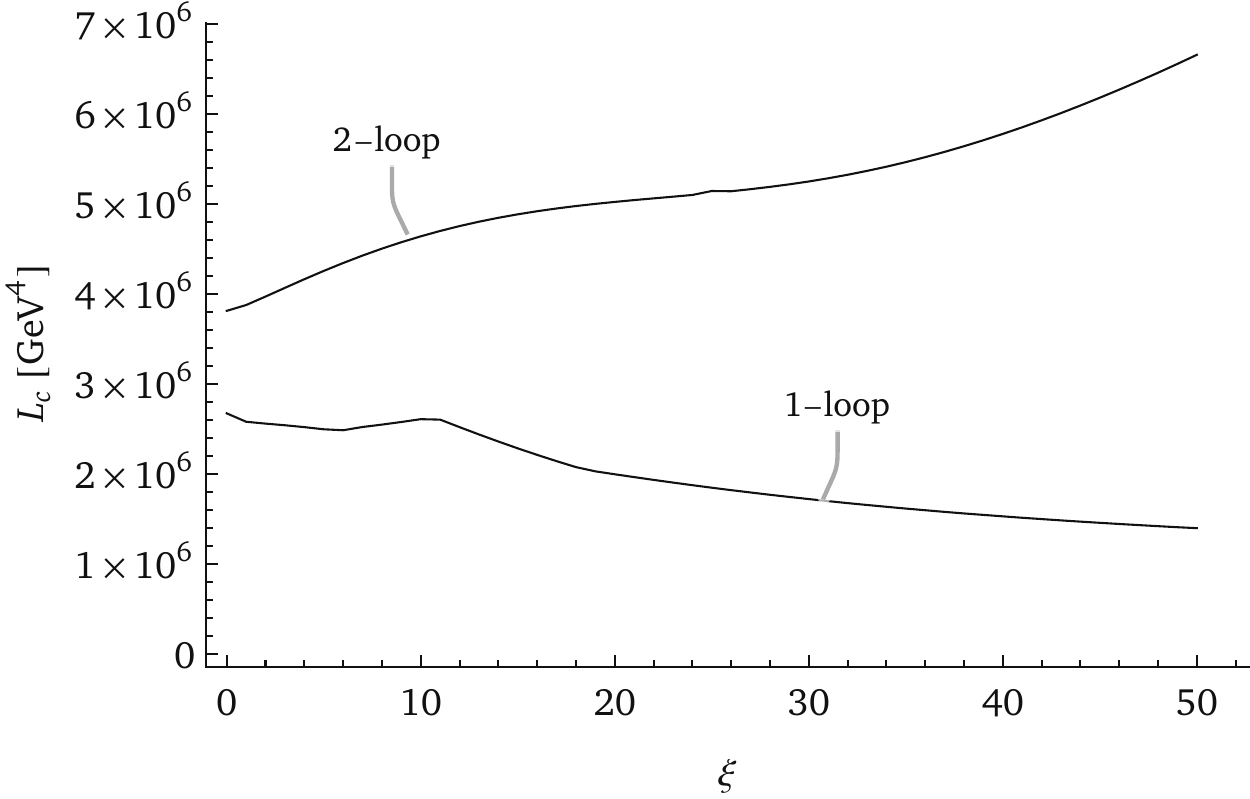}\label{sfig:TradVsXi_entr}}
	\caption{Observables at $T_c$ in the traditional method for a specific mass $m_H=45~\si{\giga\electronvolt}$, versus the gauge-fixing parameter $\xi$, of (a) the critical temperature, (b) the barrier height, (c) the phase transition strength, and (d) the latent heat.}
	\label{fig:TradVsXi}
\end{figure}

There are various observables to pick and choose from; some more gauge dependent than others.
For example, figure~\ref{sfig:TradVsXi_tc} shows that the critical temperature depends weakly on the gauge fixing parameter $\xi$. This is not unexpected, because two-loop corrections are suppressed with $e^2$ compared to the leading-order $T_c$\te likewise with 1-loop gauge dependent terms. So all gauge dependence is suppressed by $ e^2\sim 1/100$.

This is in contrast to other observables where the gauge dependence is more prominent. For example, both the barrier height in figure~\ref{sfig:TradVsXi_barr} and the transition strength in~\ref{sfig:TradVsXi_str} are quite gauge dependent,even for small $\xi$. Indeed, while the 1-loop barrier height is relatively insensitive to $\xi$\te{}the 2-loop barrier height is not. And the same story with the latent heat in figure~\ref{sfig:TradVsXi_entr}.

\subsection{Gauge-invariant method}
For the gauge-invariant method we assume that the quartic coupling scales as $\sim e^3$. All calculations are done according to section~\ref{sec:phases}.

\begin{figure}
	\centering
	\captionsetup[subfigure]{oneside,margin={0.5cm,0cm}}
	\subfloat[][]{\includegraphics[width=.47\textwidth]{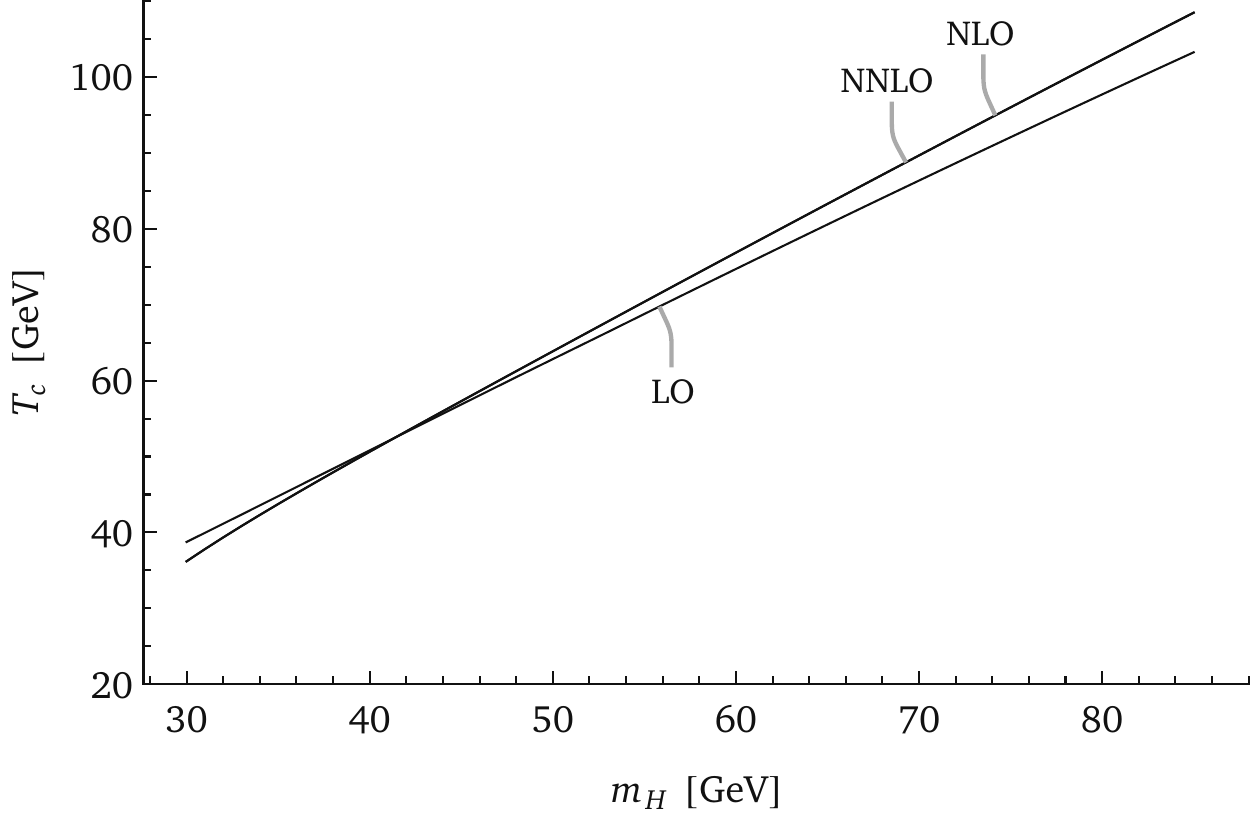}\label{sfig:GaugeIndepmH_tc}}\quad
	\subfloat[][]{\includegraphics[width=.47\textwidth]{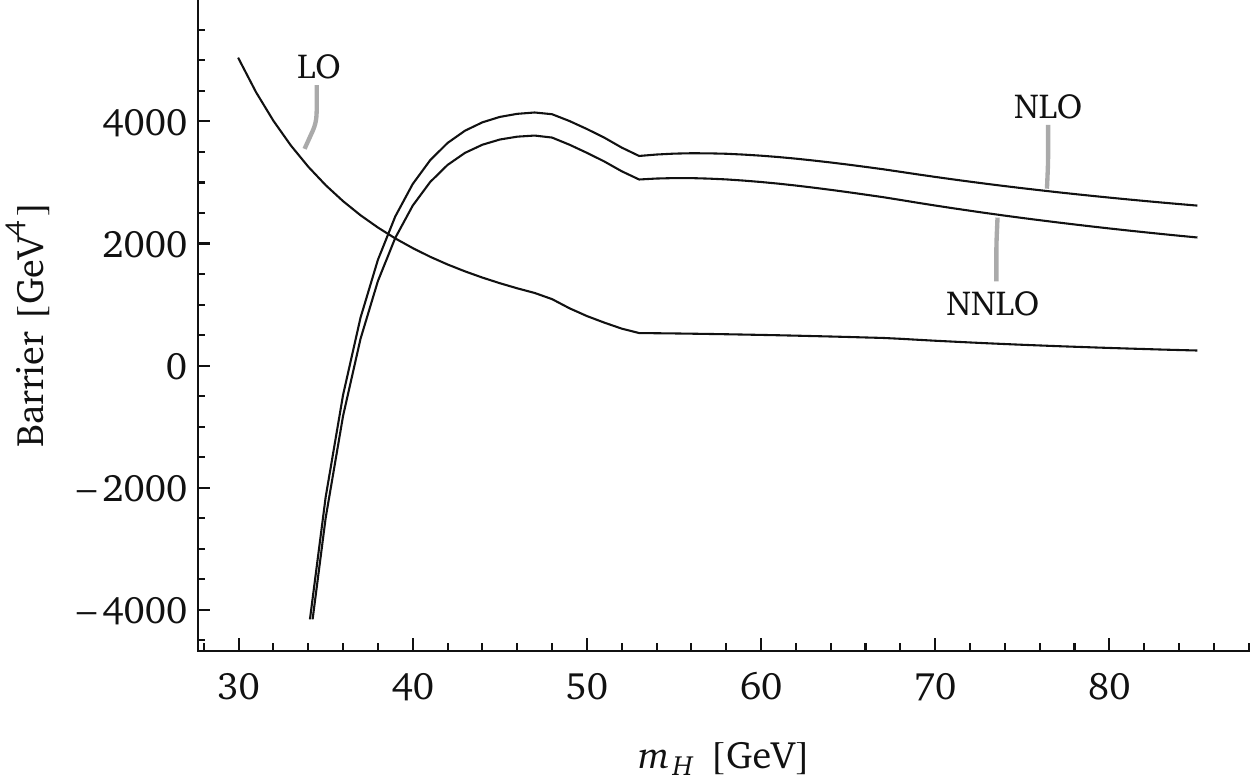}\label{sfig:GaugeIndepmH_barr}}\\
	\subfloat[][]{\includegraphics[width=.47\textwidth]{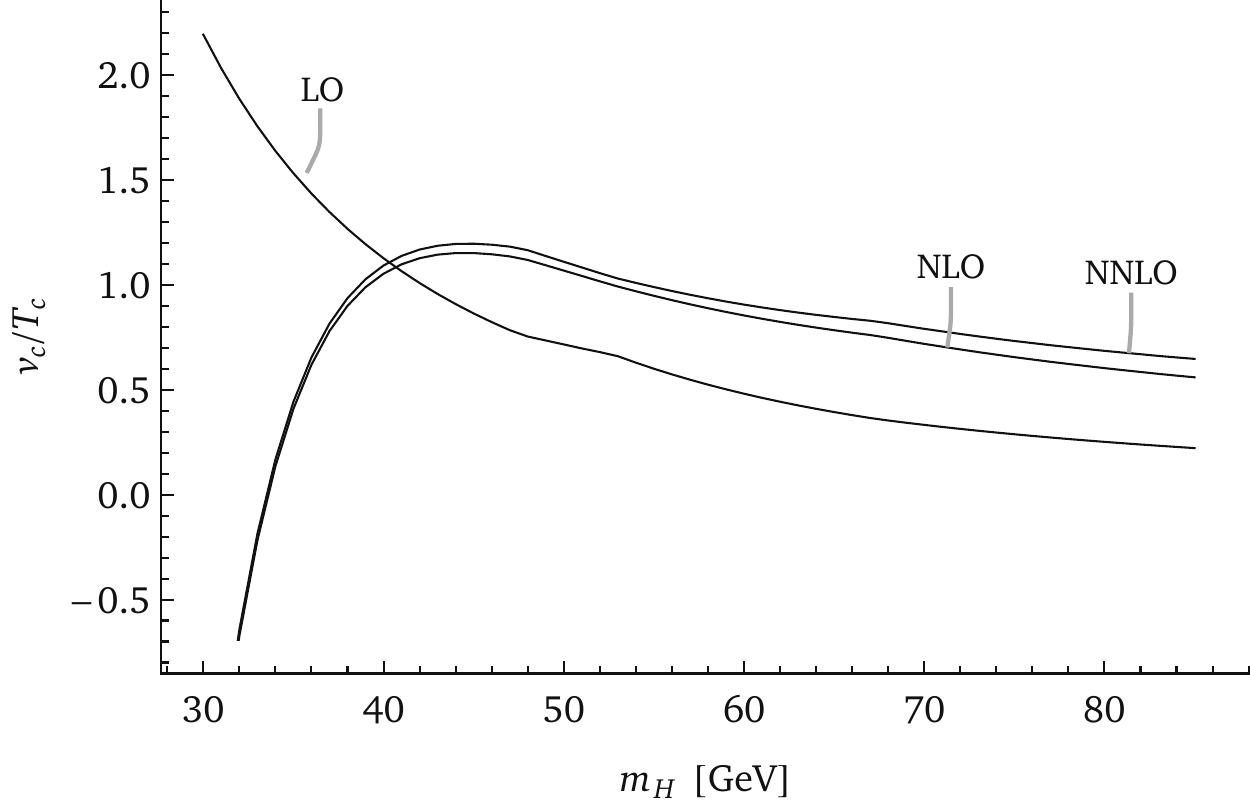}\label{sfig:GaugeIndepmH_str}}\quad
	\subfloat[][]{\includegraphics[width=.47\textwidth]{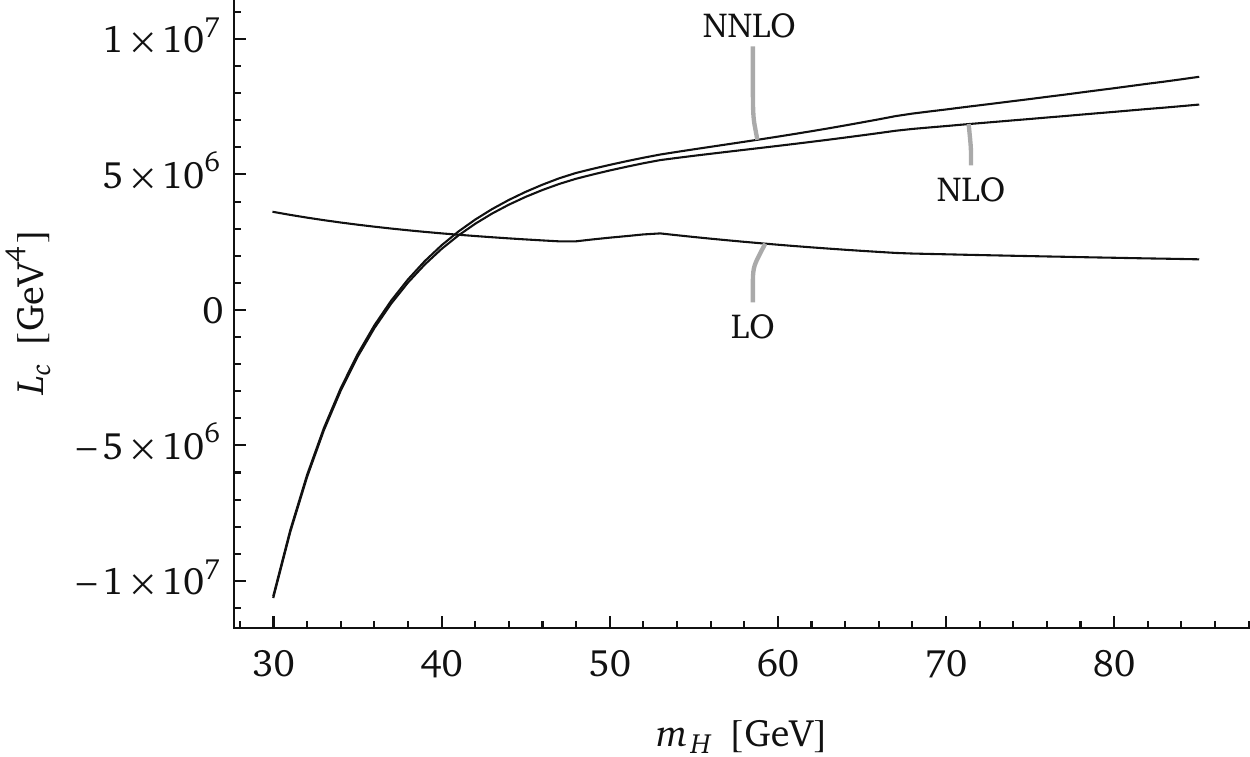}\label{sfig:GaugeIndepmH_entr}}
	\caption{Observables at $T_c$ for the gauge-invariant method. Plotted for different Higgs masses; with (a) the critical temperature, (b) the barrier height, (c) the phase transition strength, and (d) the latent heat.}
	\label{fig:GaugeIndepmH}
\end{figure}

It is evident from figure~\ref{sfig:GaugeIndepmH_tc} that higher-order corrections are suppressed when determining $T_c$. This is expected. Sub-leading corrections to $T_c$ are suppressed by a factor of $e^2 \sim 1/100$ to the leading order result.

Higher-order corrections to other observables are more pronounced. There's quite an increase of the barrier height for example. Though, the extra terms in the $\hb$ expansion are putting in some work. The extra $\TNLO\partial_T\Delta\VLO$ term in
\begin{align}
\Vbarr &= e^{-1} \Delta\VLO\big|_{\phibarr, \TLO} + e^{0}\left(\Delta\VNLO+\TNLO\partial_T\Delta\VLO\right)\Big|_{\phibarr, \TLO}\nonumber
\\&+e^{1/2} \left(\Delta\VNNLO+\TNNLO\partial_T\Delta\VLO\right)\Big|_{\phibarr, \TLO} +\mathellipsis,
\end{align}
reduces the barrier height quite a bit\te{}especially at large $m_H$.

However, note that radiative corrections to the barrier height are large both in the gauge-invariant and in the traditional method. This doesn't necessarily mean that perturbation theory is unreliable. Only that the leading-order barrier height is small. Indeed, figure~\ref{sfig:GaugeIndepmH_barr} shows that the NNLO result is of the same order as NLO. An N$^3$LO calculation would be a great cross-check on the convergence.

Finally note that all the results are unreliable for small $m_H$. Because the $\lambda\sim e^3$ scaling is not valid. So perturbation theory breaks down.
\subsection{Comparison of traditional and gauge-invariant method}
\begin{figure}
	\centering
	\captionsetup[subfigure]{oneside,margin={0.5cm,0cm}}
	\subfloat[][]{\includegraphics[width=.47\textwidth]{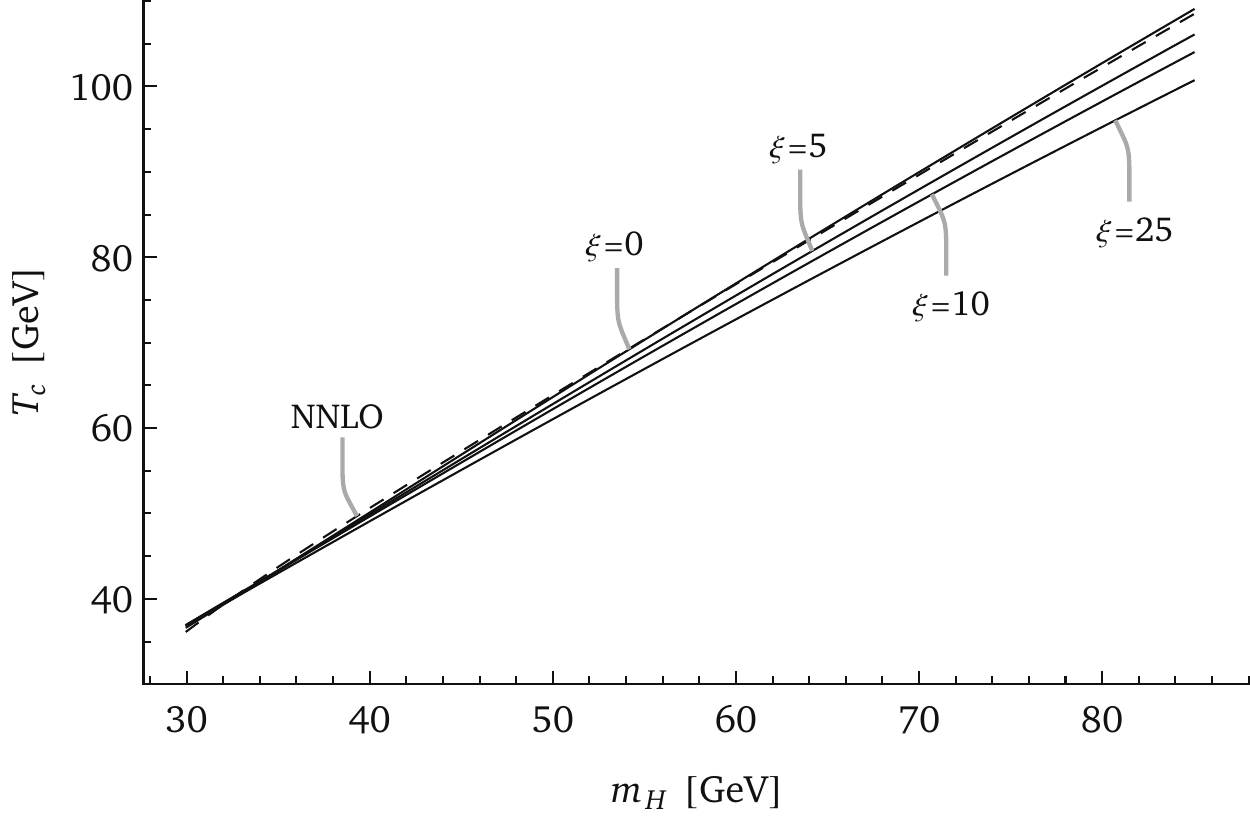}\label{sfig:TradVsGaugeIndepmH_tc}}\quad
	\subfloat[][]{\includegraphics[width=.47\textwidth]{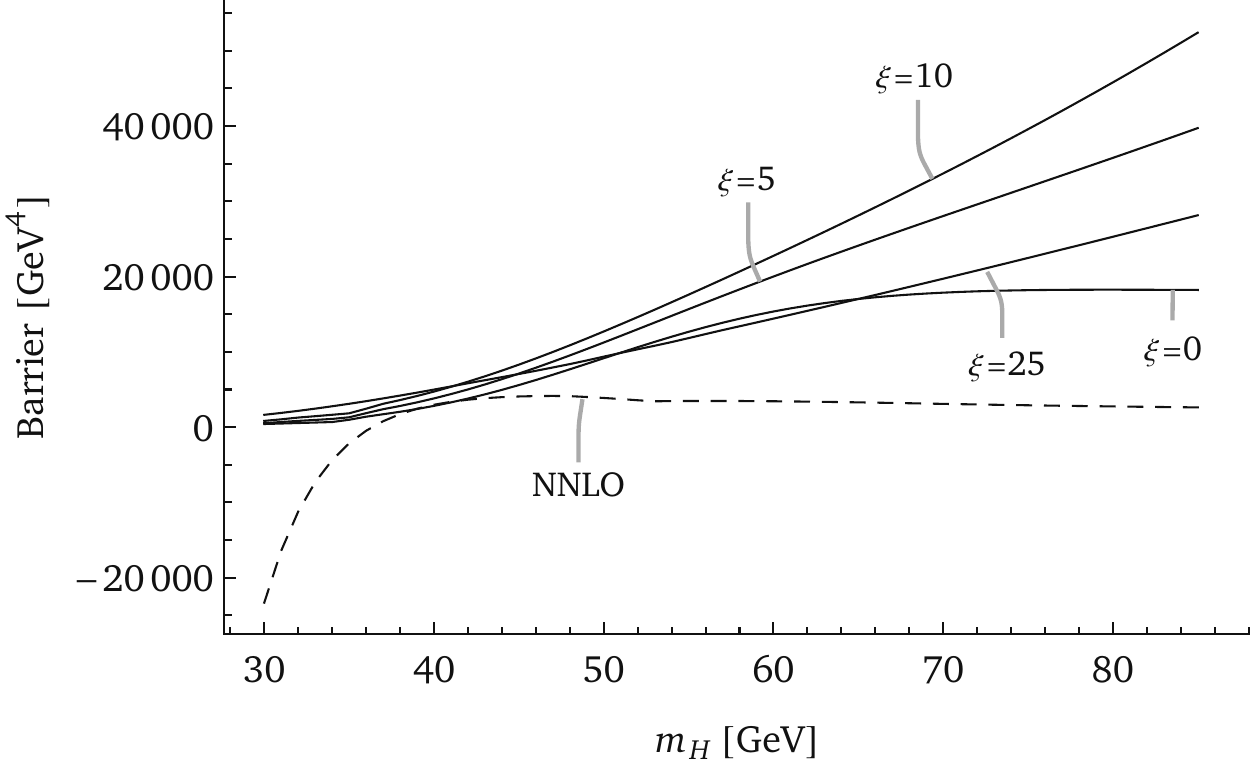}\label{sfig:TradVsGaugeIndepmH_barr}}\\
	\subfloat[][]{\includegraphics[width=.47\textwidth]{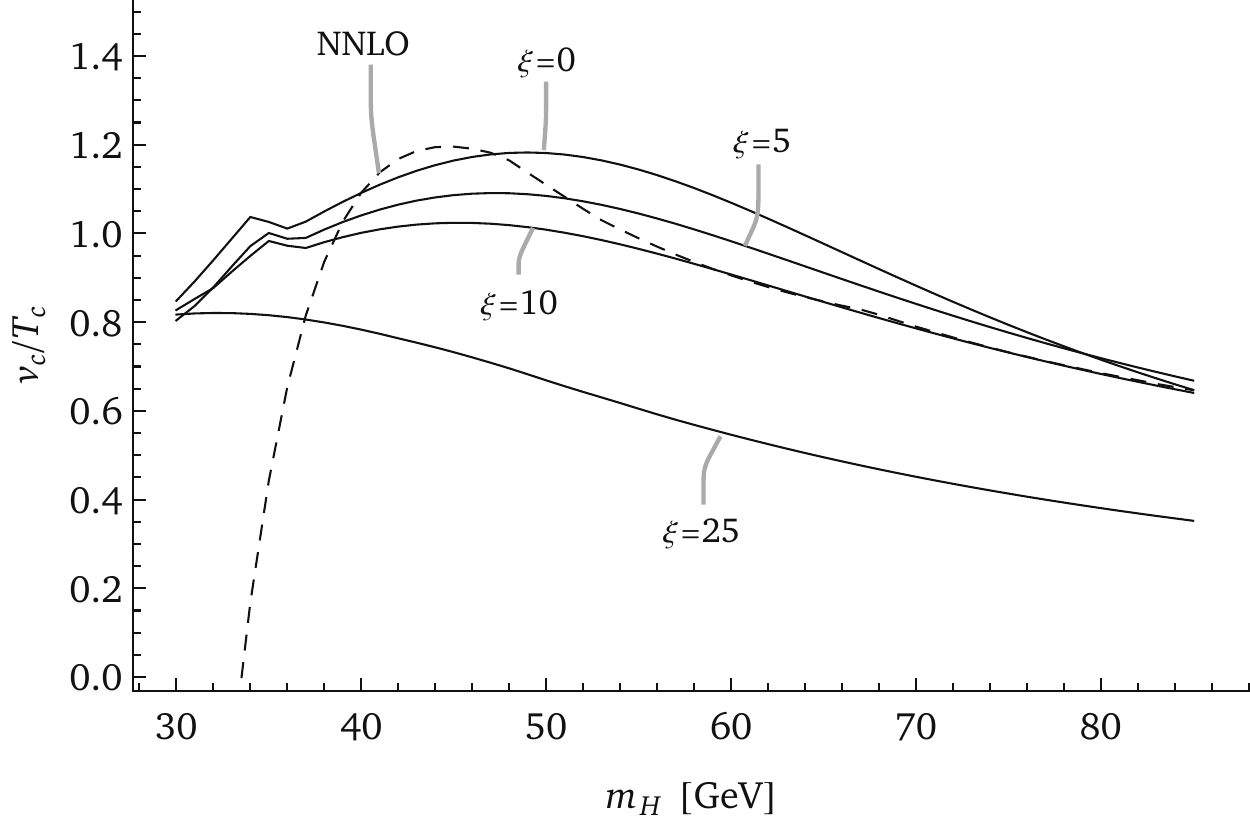}\label{sfig:TradVsGaugeIndepmH_str}}\quad
	\subfloat[][]{\includegraphics[width=.47\textwidth]{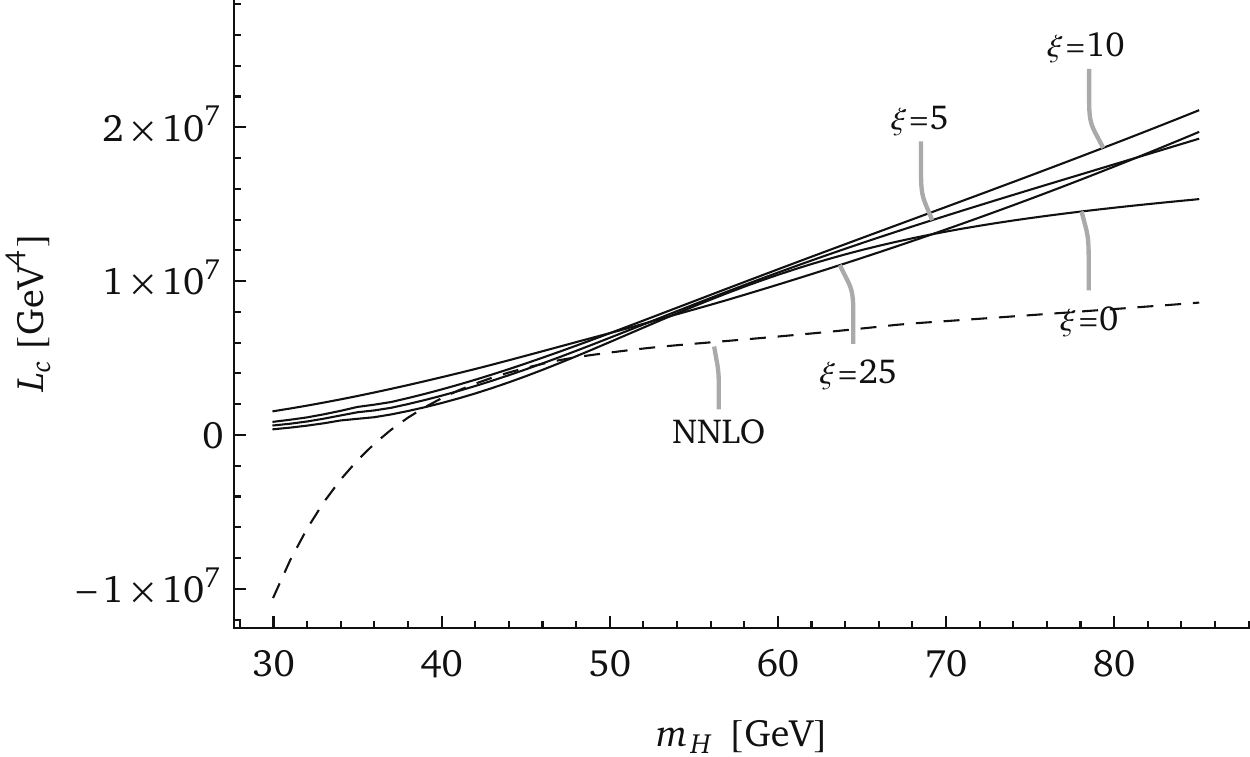}\label{sfig:TradVsGaugeIndepmH_entr}}
	\caption{Traditional method compared to the gauge-invariant method for different Higgs masses. With (a) the critical temperate, (b) the barrier height, (c) the phase transition strength, and (d) the latent heat.}
	\label{fig:TradVsGaugeIndepmH}
\end{figure}

Note from figure~\ref{sfig:TradVsGaugeIndepmH_tc} that $T_c$ is quite insensitive to $\xi$, and that the results of the gauge-invariant method coincides with that of $\xi=0$ (Landau gauge). This is however not the case for other observables. The barrier height is tremendously gauge dependent at $T_c$, as seen in figure~\ref{sfig:TradVsGaugeIndepmH_barr}. And even the Landau gauge result is an order of magnitude larger than the gauge-invariant result for certain Higgs masses. Which indicates that finite pieces, missed by the traditional method, are significant.

The phase transition strength in figure~\ref{sfig:TradVsGaugeIndepmH_str} and latent heat in~\ref{sfig:TradVsGaugeIndepmH_entr} also showcase a $\xi$ sensitivity. We conclude that all results are quite sensitive to the gauge parameter with the exception of $T_c$\te which for Landau gauge coincides with the gauge-invariant method.

%% file: tex/discussion.tex
\section{Discussion}
We showed in this paper how to include gauge-invariant resummations in the finite temperature effective potential. Beyond gauge invariant results, our method includes \textit{finite} contributions that are missed by contemporary methods.  We showed how first-order transitions, which appear highly gauge-dependent, are consistently described in this framework, and we used these methods to calculate a variety of observables\te{}comparing our results to those of the standard method.

Part and parcel of the method is the use of a consistent power-counting. Though the specific first-order scaling was first explored in~\cite{Arnold:1992rz}, a gauge-invariant method have until now remained elusive. True, some gauge invariant calculations are known~\cite{Laine:1994zq,Patel:2011th}, but these are incomplete or focus on second-order transitions.

Others~\cite{Garny:2012cg} have put bounds on the gauge dependence. These authors showed that gauge dependence is suppressed for small $\xi$. A pragmatic approach could then be to take $\xi=0$ and ignore gauge dependence all-together. Yet gauge-dependence is but the forerunner of the real issue: an inconsistent power counting. It hardly matters that the result is weakly gauge dependent when there are missing gauge-independent terms of unknown size.

In section~\ref{sec:results} we compared two different methods for calculating the critical temperature and the barrier height. The first is the vanilla gauge-dependent method, and the other is our gauge-independent method.  We conclude that only \emph{some} of the observables have small gauge dependence. Not all.

For comparison, the renormalization scale dependence might or might not be larger than the gauge dependence. But these uncertainties are fundamentally of different nature. The dependence on the renormalization scale expresses that our perturbative result is not perfect, and is reduced for each order included. Or by EFT techniques. The gauge dependence shows that terms are missing, and the problem can even worsen when further orders are included. In short, fictitious renormalization dependence is compatible with perturbative results while gauge dependence is definitely not.

There are several available avenues to continue the research in this paper. One is to calculate N$^3$LO corrections. These include $\ordo{}(T^3)$ contributions from three loops, and $\ordo{}(T)$ from two loops. The effective potential at zero temperature is known to three loops in Landau gauge~\cite{Martin:2017lqn}, so the work required would involve translating the various integral functions and master integrals to finite temperature. This calculation would also require extending the high-temperature expansion of the thermal sunset master integral to $\ordo{}(T)$\te{}see~\cite{Ekstedt:2020qyp} for a recent calculation.

Even though the step from NLO to NNLO did not induce a significant change in $T_c$, there is reason to expect that the step from NNLO to N$^3$LO will be bigger. The terms at NNLO correspond to a half-power of $e$, and they are sparse; the order N$^3$LO corresponds to a full power of $e$, with sundry diagrams. This contribution could be bigger just from combinatorics. If this is the case, this awkward pattern might continue up the ranks, where every second order in perturbation theory contributes an insignificant amount.

Another avenue: because the overarching goal of this calculation is to study extensions of the Standard Model, it would be interesting to see this method applied to other models. For models with more complicated scalar potentials some care will have to be applied in comparing the sizes of couplings. When many different couplings are involved it might be more difficult to consider the different scaling laws needed to create a barrier. But possible. With scaling laws established, then comes the issue of performing the perturbative expansion. The leading order contribution is straightforwardly calculated once the vector bosons' thermal masses are known, and is in fact easier to calculate than the traditional way (using the full 1-loop potential for all particles).

At next-to-leading order requires a two-loop calculation, which is beyond the norm of phenomenology. Because the two-loop effective potential is known for a general model~\cite{Martin:2018emo}, and since we have extended it to finite temperature in this paper, this calculation can in principle be fully automated.

Finally, thermal resummations can typically be implemented very economically using high-temperature EFT methods. It would be interesting to explore whether the power-counting of this method can be realized in such a high-temperature EFT.

\vspace*{-1em}
\section*{Acknowledgments}
We thank S.~Martin and H.~Patel for helpful discussions regarding their paper~\cite{Martin:2018emo}, and R.~Enberg for helpful comments on the manuscript. The work of A.~Ekstedt has been supported by the Grant agency of the Czech Republic, project no. 20-17490S and from the Charles University Research Center UNCE/SCI/013. This research was in part funded by the Swedish Research Council, grant no. 621-2011-5107.

%% file: tex/integrals.tex
\section{The thermal master integrals}\label{app:integrals}
In this appendix we give the leading terms in the high-temperature expansions of the thermal master integrals, using dimensional regularisation and \msbar. For readability we suppress the implicit argument $T$ to the functions.

For the 1-loop functions the expansions are known in closed form~\cite{Kapusta,Laine:2016hma}; we just include the orders we need for the calculation. The 2-loop sunset integrals $\I{x,y,z}, \IF{x,y,z}$ are given to order $T^2$.

For each integral we show its definition and its $\epsilon$ expansion. We then give the results for the leading orders in $\epsilon$ and $T$.

\subsection{Integrals for the 1-loop potential}
\subsubsection{Bosonic}
The bosonic 1-loop integral function can be written as
\begin{align}
\fB{x} & \define \frac{1}{2}T\sum_n ∫_{p} \log \left[p^2+x+(2\pi n T)^2\right].
\end{align}
The measure is $∫_{p}=(\frac{Q^2 e^\gamma}{4 \pi})^{\epsilon}\int\frac{\dif{}^{d-1} p}{(2\pi)^{d-1}}$, with $d=4-2\epsilon$ and $Q$ the \msbar{} renormalization scale. In dimensional regularisation $\fB{x}$ is separated according to
\begin{equation}
  \fB{x}=\frac{f_{-1}(x)}{\epsilon}+f(x)+\epsilon f_{\epsilon}(x)+\ordo{}(\epsilon^2),
\end{equation}
where
\begin{align}
f_{-1}(x) =& -\frac{x^2}{64 \pi^2},\\
f(x) =& -\frac{\pi^2}{90}T^4+\frac{T^2 x}{24}-\frac{T x^{3/2}}{12 \pi}-\frac{x^2}{64 \pi^2}\log\left[ \frac{e^{2\gamma_E}Q^2}{16 \pi^2 T^2} \right]+\ordo{}(x^2 \frac{x}{T^2}),\\
f_{\epsilon}(x) =& -\frac{\pi^2}{90}T^4 \left( \log\left[ \frac{Q^2}{16 \pi^2 T^2} \right]+240 \zeta'(3)+\frac{8}{3}\right)+\frac{T^2 x}{24}\log\left[ \frac{Q^2 A^{24}}{16 \pi^2 T^2} \right]\nonumber\\
&+\frac{T x^{3/2}}{12 \pi}\left(\log\left[\frac{4 x}{Q^2}\right]-\frac{8}{3}\right)+\ordo{}(x^2).\label{eq:feps}
\end{align}
The $A$ on the right hand side of equation~\eqref{eq:feps} is the Glaisher-Klinkelin constant. Its value is roughly $A \approx 1.28 $.
\subsubsection{Fermionic}
The fermionic 1-loop integral function can be written as\footnote{In this notation we are not including the "fermionic" minus sign in the basis function. This ensures congruence with~\cite{Martin:2018emo}.}
\begin{align}
\fF{x} & \define \frac{1}{2}T\sum_n ∫_{p} \log \left[p^2+x+\left(\pi (2n+1) T\right)^2\right].
\end{align}
In dimensional regularisation it is separated according to
\begin{equation}
  \fF{x}=\frac{(f_F)_{-1}(x)}{\epsilon}+f_F(x)+\epsilon (f_F)_{\epsilon}(x)+\ordo{}(\epsilon^2),
\end{equation}
where
\begin{align}
(f_F)_{-1}(x) =& -\frac{x^2}{64 \pi^2},\\
f_F(x) =& \frac{7\pi^2}{720}T^4 - \frac{T^2 x}{48}-\frac{x^2}{64 \pi^2}\log\left[\frac{e^{2\gamma_E}Q^2}{\pi^2 T^2}\right]+\ordo{}(x^2\frac{x}{T^2}),\\
(f_F)_{\epsilon}(x) =& \frac{7\pi^2}{720}T^4 \left( \log\left[ \frac{Q^2}{4^{1/7}16 \pi^2 T^2} \right]+240 \zeta'(-3)\right)\nonumber\\
&-\frac{T^2 x}{48}\log\left[ \frac{Q^2 A^{24}}{64 \pi^2 T^2} \right]+\ordo{}(x^2)\label{eq:fFeps}.
\end{align}
\subsection{The bubble}
\subsubsection{Bosonic bubble}
The bosonic bubble is
\begin{equation}
  \A{x} \define T \sum_n \int_{p} \frac{1}{p^2 + x + (2 \pi n T)^2},
\end{equation}
in dimensional regularisation we split this integral up according to
\begin{equation}
  \A{x}=\frac{A_{-1}\left(x\right)}{\epsilon}+A\left(x\right)+\epsilon A_{\epsilon}\left(x\right)+\ordo{}(\epsilon^2).
\end{equation}
In the high-temperature expansion, these individual components are given by
\begin{align}
  A_{-1}\left(x\right)&=-\frac{x}{16 \pi^2},\\
  A\left(x\right)&=\frac{T^2}{12}-\frac{T \sqrt{x}}{4 \pi}-\frac{x}{16 \pi^2}\log\left[\frac{e^{2 \gamma_E}Q^2}{16 \pi^2 T^2}\right] + \ordo{}(x\frac{x}{T^2}),\\
  A_{\epsilon}\left(x\right)&=\frac{T^2}{12} \log\left[ \frac{A^{24} Q^2}{16 \pi^2 T^2}\right] + \frac{T \sqrt{x}}{4 \pi}\left( \log\left[\frac{4 x}{Q^2}\right]-2 \right) + \ordo{}(x) \label{eq:Aeps}.
\end{align}
\subsubsection{Fermionic bubble}
The fermionic bubble is
\begin{equation}
  \AF{x} \define T \sum_n \int_{p} \frac{1}{p^2 + x + (\pi(2 n+1) T)^2},
\end{equation}
in dimensional regularisation we split this integral up according to
\begin{equation}
  \AF{x}=\frac{(A_F)_{-1}\left(x\right)}{\epsilon}+A_F\left(x\right)+\epsilon (A_F)_{\epsilon}\left(x\right)+\ordo{}(\epsilon^2).
\end{equation}
In the high-temperature expansion, these individual components are given by
\begin{align}
  (A_F)_{-1}(x)&=-\frac{x}{16 \pi^2},\\
  A_F(x)&=-\frac{T^2}{24}-\frac{x}{16 \pi^2}\log\left[\frac{e^{2 \gamma_E} Q^2}{\pi^2 T^2}\right] + \ordo{}(x\frac{x}{T^2}),\\
  (A_F)_{\epsilon}(x)&=-\frac{T^2}{24} \log\left[\frac{A^{24} Q^2}{64 \pi^2 T^2}\right] + \ordo{}(x) \label{eq:AFeps}.
\end{align}
\subsection{The double bubble}
The double bubble is not a master integral on its own, but it shows up very frequently in the integral functions, and we hence derive its finite part here. It is simply given by two bubbles multiplying each other,
\begin{equation}
  \mathbf{f}_{SS}(x, y) = \A{x}\A{y}.
\end{equation}
In dimensional regularisation we have
\begin{equation}
  \mathbf{f}_{SS}(x,y)= \frac{(f_{SS})_{-2}(x,y)}{\epsilon^2}+\frac{(f_{SS})_{-1}(x,y)}{\epsilon}+f_{SS}(x,y)+\ordo{}(\epsilon),
\end{equation}
where each of these components are given in terms of the components of $A$. We have
\begin{align}
  (f_{SS})_{-2}\left(x, y\right)&=A_{-1}(x)A_{-1}(y),\\
  (f_{SS})_{-1}\left(x, y\right)&=A(x)A_{-1}(y)+A_{-1}(x)A(y),\\
  f_{SS}\left(x, y\right)&=A(x)A(y)+ A_{\epsilon}(x)A_{-1}(y)+A_{-1}(x)A_{\epsilon}(y).
\end{align}
To be explicit, let's find the finite contribution $f_{SS}$,
\begin{align}
  f_{SS}\left(x, y\right)&=\frac{T^4}{144} -\frac{T^3}{48 \pi}\left(\sqrt{x}+\sqrt{y}\right)\nonumber\\
  &+\frac{T^2}{32 \pi^2}\left(2\sqrt{x}\sqrt{y}-\frac{1}{3}(x+y)\log\left[\frac{e^{ \gamma_E} A^{12} Q^2}{16 \pi^2 T^2}\right]\right)\nonumber\\
  &+\frac{T}{64 \pi^3}\left[x\sqrt{y}\left(\log\left[\frac{e^{ 2\gamma_E} Q^4}{64 \pi^2 T^2 y}\right]+2 \right)+y\sqrt{x}\left(\log\left[\frac{e^{ 2\gamma_E} Q^4}{64 \pi^2 T^2 x}\right]+2 \right)\right]\nonumber\\
  &+\ordo{}(x^2,x y, y^2).
\end{align}
\subsection{The sunset}
\subsubsection{Bosonic sunset}
The bosonic sunset integral is defined as
\begin{align}
  \I{x, y, z} &\define T^2 \sum_{n_p,n_q,n_l} \int_{p,q,l} \delta(p-q-l)\delta_{n_p-n_q-n_l,0}\nonumber\\
  &\times\frac{1}{p^2+ x + (2 \pi n_p T)^2}\frac{1}{q^2+ y + (2 \pi n_q T)^2}\frac{1}{l^2 + z + (2 \pi n_l T)^2}
\end{align}
In dimensional regularisation we have
\begin{equation}
    \I{x, y, z}=\frac{I_{-2}\left(x, y, z\right)}{\epsilon^2}+\frac{I_{-1}\left(x, y, z\right)}{\epsilon}+I\left(x, y, z\right)+\ordo{}(\epsilon).
\end{equation}
The infinite pieces are
\begin{align}
I_{-2}\left(x, y, z\right)&=-\frac{1}{\left(16 \pi^2\right)^2}\frac{x + y+z}{2 },\\
I_{-1}\left(x, y, z\right)&=\frac{1}{16\pi^2}\left[A(x)+A(y)+A(z)-\frac{1}{16\pi^2}\frac{x+y+z}{2}\right]\nonumber\\
&=\frac{1}{16 \pi^2}\frac{T^2}{4}+\ordo{}(T).
\end{align}
The finite piece is
\begin{align}
I\left(x, y, z\right)&=\frac{T^2}{16 \pi^2} \left(\log\left[\frac{Q}{\sqrt{x}+\sqrt{y}+\sqrt{z}}\right]+\frac{1}{2}\right)+\ordo{}(T).
\end{align}
\subsubsection{Fermionic sunset}
The fermionic sunset integrals do not contribute at order $T^2$. The reason is that for sunset integrals, the $T^2$ term arises solely due to zero modes\te{}and fermions do not have zero modes. However, we include the coefficients of the divergent terms for completeness. We use the convention that the first two masses $x,y$ (and momenta $p,q$) correspond to the fermionic modes, and $z$ (momentum $l$) corresponds to a bosonic mode; we denote the basis function $\IF{x,y,z}$, and
\begin{align}
  \IF{x,y,z} &\define T^2 \sum_{n_p,n_q,n_l} \int_{p,q,l}\delta(p-q-l)\delta_{(2n_p+1)-(2n_q+1)-n_l,0}\\
  &\times  \frac{1}{p^2+ x + (\pi (2 n_p+1) T)^2}\frac{1}{q^2+ y + (\pi (2 n_q+1) T)^2}\frac{1}{l^2 + z + (2\pi n_l T)^2}.\nonumber
\end{align}
First, let's note as usual that in dimensional regularisation we have
\begin{equation}
  \IF{x,y,z}=\frac{(I_{F})_{-2}\left(x, y, z\right)}{\epsilon^2} + \frac{(I_{F})_{-1}\left(x, y, z\right)}{\epsilon}+ I_{F}\left(x, y, z\right)+\ordo{}(\epsilon).
\end{equation}
The infinite pieces are
\begin{align}
(I_{F})_{-2}\left(x, y, z\right)&=-\frac{1}{\left(16 \pi^2\right)^2}\frac{x + y+z}{2},\\
(I_{F})_{-1}\left(x, y, z\right)&=\frac{1}{16 \pi^2}\left(A_{F}(x)+A_{F}(y)+A(z)-\frac{1}{16 \pi^2}\frac{x+y+z}{2}\right)\nonumber\\
&=\ordo{}(T \sqrt{z}).
\end{align}
The finite piece is
\begin{equation}
(I_{F})\left(x, y, z\right) = \ordo{}(T \sqrt{z}).
\end{equation}
\section{Thermal Integral Functions}\label{app:intfuncs}
The integral functions used by Martin and Patel in~\cite{Martin:2018emo} capture concisely the Lorentz structure of the different classes of diagrams. We use them in our calculations, but there is an added complication at finite temperature. The bubble evaluated at zero, $\A{0}=\frac{T^2}{12}$, gives a zero contribution at zero temperature, but a nonzero one at finite temperature. This term generally arises from partial-fraction decompositions performed to calculate integral functions that involve vector propagators.

We have recalculated the integral functions and retained these extra terms. Below we show the additional terms that are missing from the equations in section~III.B of~\cite{Martin:2018emo}; we hide the previously given terms behind "\ldots":
\begingroup
  \allowdisplaybreaks
\begin{align}
  \mathbf{f}_{SSV}(x,y,z)=&\mathellipsis +\frac{1}{z}\left(x-y\right)\A{0}\left(\A{x}-\A{y}\right),\\
  \mathbf{f}_{SS\overline{V}}(x,y,z)=&\mathellipsis -\frac{1}{z}\left(x-y\right)\A{0}\left(\A{x}-\A{y}\right),\\
  \mathbf{f}_{\overline{VV}S}(x,y,z)=&\mathellipsis +\frac{1}{4 x y}\A{0}\Big{[}\left(x-z\right)\A{x}+\left(y-z\right)\A{y}-\left(x+y\right)\A{z}\nonumber\\
  &\hspace{6.5em} +z\A{0}\Big{]},\\
  \mathbf{f}_{GS\overline{V}}(x,y,z)=&\mathellipsis +\frac{1}{2z}\left(x-y\right)\A{0}\left(\A{x}-\A{y}\right),\\
  \mathbf{f}_{VGG}(x,y,z)=&\mathellipsis-
  \frac{1}{4 x}\left(y-z\right)\A{0}\left(\A{y}-\A{z}\right),\\
  \mathbf{f}_{\overline{V}GG}(x,y,z)=&\mathellipsis+
  \frac{1}{4 x}\left(y-z\right)\A{0}\left(\A{y}-\A{z}\right),\\
  \mathbf{f}_{VVG}(x,y,z)=&\mathellipsis+
  \frac{1}{4 x y}\A{0}\Big{[}x \left(z-x\right)\A{x}-y\left(z-y\right)\A{y}\nonumber\\
  &\hspace{6.5em}+\left(x-y\right) \left(x+y-z\right)\A{z}\Big{]},\\
  \mathbf{f}_{VVV}(x,y,z)=&\mathellipsis+\frac{1}{4 x y z}\A{0}
  \Big{\{}\left(-y^3-z^3+(4 d-7)\left[x y^2 + x z^2 - x^2 y - x^2 z\right] \right)\A{x}\nonumber\\
  &\hspace{6.6em}+\left(-x^3-z^3+(4 d-7)\left[y x^2 + y z^2 + y^2 x + y^2 z\right] \right)\A{y}\nonumber\\
  &\hspace{6.6em}+\left(-x^3-y^3+(4 d-7)\left[z x^2 + z y^2 + z^2 x + z^2 y\right] \right)\A{z}\nonumber\\
  &\hspace{6.6em}+\left(x^3+y^3+z^3\right)\A{0}\Big{\}},\\
  \mathbf{f}_{\overline{V}VV}(x,y,z)=&\mathellipsis+\frac{1}{4 x y z}\A{0}
  \Big{\{}\left(-x \left(y^2+z^2\right)+y^3+z^3\right)\A{x}\nonumber\\
  &\hspace{6.6em}+\left(z^3 + x y^2 +yz\left(4d-7\right)\left[z-y\right]\right)\A{y}\nonumber\\
  &\hspace{6.6em}+\left(y^3 + x z^2 +yz\left(4d-7\right)\left[y-z\right]\right)\A{z}\nonumber\\
  &\hspace{6.6em}-\left(y^3+z^3\right)\A{0}\Big{\}},\\
  \mathbf{f}_{\overline{VV}V}(x,y,z)=&\mathellipsis -\frac{1}{4 x y}z\A{0}\Big{[}\left(z-x\right)\A{x}+\left(z-y\right)\A{y}+\left(x+y\right)\A{z}\nonumber\\
  &\hspace{6.5em}-z \A{0}\Big{]},\\
  \mathbf{f}_{\eta\eta V}(x,y,z)=&\mathellipsis-\frac{1}{2 z}\left(x-y\right)\mathbf{A}(0)\left(\A{x}-\A{y}\right),\\
  \mathbf{f}_{\eta\eta \overline{V}}(x,y,z)=&\mathellipsis+\frac{1}{2 z}\left(x-y\right)\mathbf{A}(0)\left(\A{x}-\A{y}\right),\\
  \mathbf{f}_{FF V}(x,y,z)=&\mathellipsis-\frac{1}{z}(x-y)\A{0}\left(\AF{x}-\AF{y}\right),\\
  \mathbf{f}_{FF \overline{V}}(x,y,z)=&\mathellipsis+\frac{1}{z}(x-y)\A{0}\left(\AF{x}-\AF{y}\right).
\end{align}
\endgroup
All of the integral functions above have a nice consistency check: all limits of masses going to zero should be safe. At finite temperature the right results are only obtained if the above formulas are used.

\section{Resummations and Vector Bosons}\label{app:resum}
\subsection{Resummation of Longitudinal Gauge Boson Masses}\label{app:longitudinal}
In effective potential calculations we need all four modes of the gauge boson. For a gauge boson carrying four-momentum $k^\mu$, we classify these as one longitudinal mode corresponding to fluctuations along the direction spanned by $k^\mu$, and three transverse modes. The longitudinal modes are unphysical and their contributions must cancel against other unphysical degrees of freedom. In our Fermi gauge calculations this manifests as cancellations between contributions from ghosts, longitudinal modes, and Goldstones. The three remaining transverse modes correspond to physical degrees of freedom.

At finite temperature the Lorentz invariance is broken from four to three dimensions. The three transverse modes discussed above now further split into one mode corresponding to fluctuations along the direction $\vec{k}$. A \emph{3D-longitudinal} mode, and two transverse modes.

This distinction is important because at finite temperature only the $3$D-longitudinal modes should be resummed. At one loop this is straightforward, as the three transverse modes contribute independently. For a gauge boson with squared mass $X$ and resummed squared mass $X_L$, the resummation is performed as
\begin{equation}
  -3\frac{T}{12 \pi} X^{3/2} \rightarrow -2\frac{T}{12 \pi} X^{3/2}-\frac{T}{12 \pi} X_L^{3/2}.
\end{equation}

At two loops and higher the situation is more complicated, because interactions will intermingle the modes. The integral functions used in~\cite{Martin:2018emo} capture the Lorentz structure of the different diagrams; to perform this resummation we will need to project out the contributions of the different modes.

In that vein, consider the propagator of a massive vector boson,
\begin{equation}
   D^{\mu\nu}(k)=A(k^2) P^{\mu\nu}(k)+B (k^2) \frac{k^\mu k^\nu}{k^2},
 \end{equation}
where we hid masses and gauge-fixing parameters in the Lorentz invariant functions $A$ and $B$. We can focus on the projection operators: $\smash{P^{\mu\nu}(k)\define g^{\mu\nu}-\frac{k^\mu k^\nu}{k^2}}$ projects onto the three transverse modes; $\frac{k^\mu k^\nu}{k^2}$ projects onto the longitudinal mode. To find the individual contributions from the transverse modes we introduce the $3$D-longitudinal projector $P_L^{\mu\nu}$ and the corresponding transverse projector $P_T^{\mu\nu}$, such that
\begin{equation}
  P^{\mu\nu}=P_L^{\mu\nu}+P_T^{\mu\nu}.
\end{equation}
Using the properties of these projectors~\cite{BUCHMULLER1993387}, we can derive the contributions from the $3$D-longitudinal modes to the various integral functions. We use the notation that an index $V^L$ corresponds to a $3$D-longitudinal mode, and an index $V^T$ for the transverse modes. The non-zero integral functions that include at least one $3$D-longitudinal mode are the following:
\begin{align}
  \mathbf{f}_{V^L S}(x,y)&=\A{x}\A{y},\\
  \mathbf{f}_{V^L V^L S}(x,y,z)&=-\I{x,y,z},\\
  \mathbf{f}_{V^L V^L G}(x,y,z)&=(x-y)\I{x,y,z} + \left(\A{x}-\A{y}\right)\A{z},\\
  \mathbf{f}_{V^T V^L}(x,y)&=(d-2)\A{x}\A{y},\\
  \mathbf{f}_{\overline{V} V^L}(x,y)&=\A{x}\A{y},\\
  \mathbf{f}_{V^T V^L V^L}(x,y,z)&=\frac{1}{x}\Big{[}-\lambda(x,y,z)\I{x,y,z}+(y-z)^2\I{0,y,z}\nonumber\\
  &\hspace{2.em} + (z-x-y)\A{x}\A{y} + (y-x-z)\A{x}\A{z}\nonumber\\
  &\hspace{2.5em} + x \A{y}\A{z} +(y-z)(\A{y}-\A{z})\mathbf{A}(0)\Big{]},\\
  \mathbf{f}_{\overline{V} V^L V^L}(x,y,z)&=\frac{1}{x}\Big{[}(y-z)^2(\I{x,y,z}-\I{0,y,z})+(y-x-z)\A{x}\A{y} \nonumber\\
  &\hspace{2.5em} +(z-x-y)\A{x}\A{z}+(z-y)(\A{y}-\A{z})\A{0}\Big{]},\\
  \mathbf{f}_{FF V^L}(x,y,z)&=(x+y-z)\IF{x,y,z}+\AF{x}\AF{y}-\left(\AF{y}+\AF{x}\right)\A{z},\\
  \mathbf{f}_{\overline{F}\overline{F} V^L}(x,y,z)&=  2 \IF{x,y,z}.
\end{align}
In the above list we did not include permutations. As an example, there is also the integral function $\mathbf{f}_{V^L V^T}(x,y)$. For each of the integral functions above there is also a corresponding one that only includes the transverse modes. We can find them by using the full result together with the formulas above. For example
\begin{equation}
  \mathbf{f}_{VV}(x,y)=\mathbf{f}_{V^T V^T}(x,y)+\mathbf{f}_{V^T V^L}(x,y)+\mathbf{f}_{V^L V^T}(x,y).
\end{equation}

Furthermore, only the vectors' zero-modes should be resummed, and we need to pick out their contributions to the master integrals. These are
\begin{align}
  \left.\fB{x}\right|_{0}&=-\frac{T x^{3/2}}{12 \pi},\\
  \left.\A{x}\right|_{0}&=-\frac{T \sqrt{x}}{4 \pi},\\
  \left.\I{x,y,z}\right|_{0}&=\frac{T^2}{16 \pi^2} \left(\log\left[\frac{Q}{\sqrt{x}+\sqrt{y}+\sqrt{z}}\right]+\frac{1}{2}\right)+\ordo{}(T),\\
  \left.\IF{x,y,z}\right|_{0}&=\ordo{}(T\sqrt{z}).
\end{align}
\subsection{Thermal Gauge Boson Masses in the Standard Model}
In the Standard Model there are additional complications.
At finite temperature neutral $3$D-longitudinal modes $W^3_L, B_L$ have the mass matrix
\begin{equation}
\begin{pmatrix}
W^3_L & B_L
\end{pmatrix}
\begin{pmatrix}
\frac{1}{4}g^2 \phi^2 + \Pi_W(T)  & -\frac{1}{4} g g'\phi^2 \\
-\frac{1}{4} g g'\phi^2 & \frac{1}{4} g'^2 \phi^2 +\Pi_B(T)
\end{pmatrix}
\begin{pmatrix}
W^3_L \\
B_L
\end{pmatrix},
\end{equation}
where the thermal self-energies are~\cite{Kapusta}
\begin{align}
&\Pi_W(T)=\frac{11}{6}g^2T^2,
\\& \Pi_B(T)=\frac{11}{6}g'^2 T^2.
\end{align}

This mass matrix is diagonalized by an angle $\theta'$\te depending implicitly on $T$ and $\phi$. The explicit form of $\theta'$ can be found using the mass matrix above, we neglect to show it here and simply note the limits
\begin{align}
 \phi>0, ~T\rightarrow 0 &\implies \theta' \rightarrow \theta_W,\\
 \phi\rightarrow 0 &\implies \theta' \rightarrow 0.
\end{align}

Massive eigenstates are $Z_L,A_L$; they have squared masses
\begin{align}
  Z_L~/~A_L = \frac{1}{2}&\bigg(\Pi_W+\Pi_B+Z\nonumber
  \\&\hspace{2em}\pm\left.\sqrt{\left(\Pi_W+\Pi_B+Z\right)^2-\left(4 \Pi_W \Pi_B+ \Pi_W g'^2 \phi^2+\Pi_B g^2 \phi^2\right)}\right),
\end{align}
where we mean that $Z_L$ has a $+$ sign in front of the square root, and $A_L$ has a $-$ sign. These
masses behave as expected,
\begin{align}
 T \rightarrow 0 &\implies Z_L\rightarrow Z,~ A_L \rightarrow 0,\\
 \phi \rightarrow 0 & \implies Z_L\rightarrow \Pi_W(T),~ A_L\rightarrow \Pi_B(T).
\end{align}

The final complication is that resummed $3$D-longitudinal modes have different coupling constants; the tensors given in~\cite{Martin:2018emo} must be modified. Using the notation $c_\theta\define \cos\theta,s_\theta\define \sin\theta$, and letting $I_f$ denote weak isospin of left-handed fermion $f$, $Y_f$ hyper-charge, and $Q_f$ electric charge,
\begin{alignat}{2}
g^{Z_L f}_{f}&=I_f g \ct-Y_f g' \st, &&
\\ g^{Z_L  \bar{f}}_{\bar{f}}&=Q_f g' \st, &&
\\ g^{A_L  u}_{u}&=\frac{1}{6}\left(g' \ct+3 g \st\right),\hspace{2em} g^{A_L  \bar{u}}_{\bar{u}}&&=-\frac{2}{3}\ct g',
\\ g^{A_L  d}_{d}&=\frac{1}{6}\left(g'\ct-3 g\st\right), \hspace{2em} g^{A_L  \bar{d}}_{\bar{d}}&&=\frac{1}{3}\ct g',
\\ g^{A_L  l^{-}}_{l^{-}}&=-\frac{1}{2}\left(g' \ct+g\st\right), \hspace{2em} g^{A_L  l^{+}}_{l^{+}}&&=\ct g',
\\ g^{A_L  \nu}_{\nu}&=-\frac{1}{2}\left(g' \ct-g \st\right).&&
\end{alignat}

\begin{alignat}{2}
g^{Z_L }_{G_0 H}&=\frac{1}{2}\left(g \ct+g'\st\right), \hspace{2em} g^{Z_L }_{G_I G_R}&&=\frac{1}{2}\left(g \ct-g'\st\right),
\\g^{A_L }_{G_I G_R}&=\frac{1}{2}\left(g' \ct+g \st\right), \hspace{2em} g^{A_L }_{G_0 H}&&=-\frac{1}{2}\left(g' \ct-g\st\right).
\end{alignat}

\begin{alignat}{2}
G^{Z_L  Z_L }_{H}&=\frac{1}{2}\left(g \ct+g' \st\right)^2 \phi, &&
\\G^{Z_L  W_R}_{G_R}&=-\frac{1}{2}g g' \st \phi, \hspace{2em} &&G^{Z_L  W_I}_{G_I}=\frac{1}{2}g g' \st \phi,
\\G^{A_L  W_R}_{G_R}&=\frac{1}{2}g g' \ct \phi, \hspace{2em} && G^{A_L  W_I}_{G_I}=-\frac{1}{2}g g' \ct\phi,
\\G^{A_L  A_L }_{H} &=\frac{1}{2}\left(g' \ct-g \st\right)^2. &&
\end{alignat}

\begin{equation}
g^{Z_L  W_R W_I}=g \ct, \hspace{2em} g^{A_L  W_R W_I}=g \st.
\end{equation}

Any coupling left out of this list is identical to that given in~\cite{Martin:2018emo}.